\documentclass[11pt,a4paper]{article}

\usepackage[a4paper,
            left=2.2cm,
            right=2.2cm,
            top=2.5cm,
            bottom=2.5cm]{geometry}

\usepackage{amsmath,amssymb,amsfonts}
\usepackage{graphicx}
\usepackage{subfig}
\usepackage{booktabs,multirow}
\usepackage{xcolor}
\usepackage{physics}
\usepackage{cite}
\usepackage{comment}
\usepackage{subcaption}
\usepackage[hidelinks]{hyperref}
\usepackage{url}

\begin{document}

\title{Full-Wave Optical Modeling of Leaf Internal Light Scattering Dynamics
with Potential Applications for Early Detection of Foliar Fungal Disease}

\author{
Da-Young Lee$^{1}$ \and
Dong-Yeop Na$^{1}$\thanks{Corresponding author: dyna22@postech.ac.kr}
\\
\small $^{1}$Department of Electrical Engineering, Pohang University of Science and Technology (POSTECH), \\Pohang, 37673, Republic of Korea
}

\date{}

\maketitle   

\begin{abstract}
Light that interacts with plant leaves undergoes reflection, transmission, scattering, and absorption that collectively contribute to leaf optical properties. 
Modifications in leaf architecture disrupt these optical properties, including light-scattering dynamics within the leaf, which in turn, affect leaf photosynthetic performance. 
Previous studies on internal leaf light scattering have relied primarily on ray-tracing approaches (e.g., Raytran) or radiative-transfer formalisms (e.g., PROSPECT). 
These methods are based on high-frequency approximations and therefore cannot account for diffraction and coherent multiple scattering in wavelength-scale leaf tissue geometries, in contrast to full-wave electromagnetic simulations.
Here, we present the use of Finite-Difference Time-Domain (FDTD) accelerated on Graphics Processing Units (GPUs) to simulate internal light-scattering dynamics using representative dicot and monocot leaf cross-section image geometries. 
Microscopy images were segmented and assigned wavelength-dependent complex refractive indices. 
Our results accurately reproduced the well-known reflectance and transmittance characteristics of healthy leaves, through Lin's concordance analysis showing showing high ($C_b = 0.90$) and moderate concordance ($C_b = 0.79$) with respect to the PROSPECT model (reference) for dicot and monocot leaves, respectively.
Furthermore, an early-stage, fungal pathogen infection, mimicking melanized hyphae penetrating the cuticle and upper epidermis, was also simulated. 
Our results showed that diseased leaves exhibited a pronounced decrease in visible green reflectance and a marked suppression of the NIR reflectance plateau, in agreement with experimental observations. Minor discrepancies in the visible-band reflectance are expected to be mitigated through more advanced geometric and material modeling.
Our proof-of-concept study is the first-ever attempt to use a FDTD full-wave optical modeling approach for plant-leaf light-scattering analysis, presenting a light-scattering-based optical modeling framework to simulate internal light scattering within plant leaves pre- and post-penetration of leaf tissue during early fungal infection. 
The findings in this study will serve as a foundation for obtaining insights into the use of light scattering as an indicator for pre-symptomatic plant fungal disease detection.
\end{abstract}

\section{Introduction}
\subsection{General introduction}
Plant pathogens cause diseases that present serious challenges in optimal crop production~\cite{strange2005plant,mengiste2025contrasting}. 
Biotic stressors detrimentally impact food production~\cite{strange2005plant}, and fungal plant pathogens have been reported to the predominant group of pathogens contributing to global crop yield loss~\cite{oerke2006crop, strange2005plant, dean2012top, savary2019global}. 
A significant decrease in food crop production poses a threat not only in terms of crop yield and economic loss, but also decreases species diversity, increases mitigation costs for disease management, and downstream effects on human welfare~\cite{bebber2015crop}. 
Hence, novel physics-based approaches capable of early and reliable disease detection are essential for mitigating the severe agricultural, ecological, and socio-economic impacts of fungal plant pathogens.

Plant leaves are the main surfaces of plant vegetation and have evolved to harvest the light required for effective photosynthesis to occur~\cite{vogelmann2014leaf}. 
This photosynthetic organ contains chloroplasts, whose light exposure maximizes photosynthesis~\cite{gotoh2018palisade,terashima2011leaf} and the heterogeneous micro-light environment within the leaf influences photosynthetic efficiency~\cite{karabourniotis2021optical}.
Plant leaves represent a particularly complex scattering environment.
Light--tissue interactions are governed by cellular geometry, refractive-index discontinuities at air--tissue interfaces, and the spatial distribution of pigments and water~\cite{vogelmann1994light}. 
As light penetrates the leaf, it undergoes repeated refraction at cell walls, scattering within intercellular air spaces, and absorption by chromophores~\cite{jacquemoud2008modeling}. 
These sequential processes reshape the internal light field as it propagates through the epidermis, the palisade mesophyll, and the spongy mesophyll, collectively determining the leaf’s reflectance, transmittance, and absorption spectra.
Understanding leaf optical behavior ultimately requires the fundamental physics of light--matter interaction:
when an electromagnetic (EM) wave (or light) encounters an inhomogeneous medium, defined both by geometric complexity and refractive-index variation, its propagation is modified through reflection, refraction, scattering, absorption, and diffraction~\cite{jackson2012classical,chew1999waves,balanis2012advanced}.

Because optical signatures of leaves directly reflect their underlying microstructure and biochemical composition, hyperspectral imaging has emerged as an effective non-invasive tool for monitoring plant health and diagnosing physiological stress or early infection. 
Healthy leaves exhibit scattering dominated by refractive-index contrasts and well-organized cellular structures, whereas early infection induces microstructural changes---cell collapse, pigment degradation, and fungal melanin deposition~\cite{szechynska2025application,zarco2018previsual}---that perturb scattering paths and modify local absorption. 
Accurate modeling of these processes clarifies how subtle anatomical disruptions generate measurable shifts across the visible (VIS) and near-infrared (NIR) spectrum, which are routinely captured by hyperspectral imaging systems.

\subsection{Previous approaches for modeling light scattering within leaf}
A wide range of studies have been devoted to modeling light reflectance and transmittance in plant leaves. 
The foundational work by Allen, Richardson, and Gausman~\cite{allen1968interaction,allen1969interaction,gausman1970relation,allen1970mean,gausman1971age} established the physical basis of leaf optics by quantifying effective optical constants and relating reflectance to leaf maturity, water content, and thickness. 
Kumar and Silva~\cite{kumar1973reflectance,kumar1973light} later introduced geometric ray-tracing approaches to model internal reflections within leaf cross-sections under simplified optics. 
Tucker and collaborators~\cite{tucker1977leaf,tucker1977spectral,tucker1980remote} developed stochastic radiative models that treated leaf tissues as probabilistic scattering media, enabling canopy-scale reflectance estimation.

As the field progressed, statistical and Monte Carlo approaches were developed to capture more realistic tissue heterogeneity. 
Stochastic models~\cite{tucker1977leaf,maier1999slop}, Raytran~\cite{govaerts1996three,govaerts2002raytran}, and meshed Monte Carlo methods such as MMC-fpf~\cite{watte2015modeling} facilitated 3D photon-transport simulations, enabling the analysis of bidirectional reflectance distribution function (BRDF), scattering phase functions, and complex mesophyll architectures.

Notably, the PROSPECT family of models~\cite{jacquemoud1996estimating,feret2008prospect,feret2017prospect,feret2021prospect,yu2023piosl,su2024monitoring,feret2024prospect,du2025prospect} has been the most influential framework for leaf-optics modeling.
These models treat a leaf as a multilayer effective medium linking biochemical constituents---pigments, water, proteins, and carbon-based compounds---to macroscopic reflectance and transmittance spectra. 
Over three decades, PROSPECT variants have expanded to incorporate pigment partitioning, structural variations, lifecycle dynamics, nitrogen absorption, and fluorescence corrections, making them the standard for remote sensing and biochemical inversion.

Beyond biochemical inversion, several studies underscored the importance of detailed anatomy in shaping internal light environments. 
Epidermal microlensing~\cite{bone1985epidermal}, anatomical radiative-transfer models~\cite{govaerts1996three}, photon-transport simulations~\cite{ustin2001simulation}, and ray-tracing of realistic leaf tissues~\cite{xiao2016influence} demonstrated the strong influence of cellular geometry on irradiance distribution and photosynthetic efficiency. 
Recent work further highlighted spectral filtering by layered mesophyll structures~\cite{xu2023light}, the optical role of palisade geometry~\cite{borsuk2025palisade}, and oblique-illumination enhancement~\cite{nikolopoulos2024leaf}. 
Polarimetric modeling~\cite{kallel2018leaf,he2025multispectral} and chlorophyll-fluorescence studies~\cite{pedros2008chlorophyll} extended these frameworks to vector and emission domains.

\subsection{Fundamental limitations of existing models}
Despite these advances, a common limitation remains: most previous models propagate photons using ray- or radiative-transfer (RT)-based assumptions.
Such methods inherently rely on the high-frequency approximation, which assumes that scattering structures are much larger than the wavelength and that light travels as independent rays undergoing reflection, refraction, and absorption. 
This assumption greatly reduces computational cost and has enabled decades of progress, but it prevents these models from capturing coherent EM phenomena such as interference, diffraction, near-field coupling, guided-mode formation, and resonant multiple scattering.

This limitation becomes critical because many internal components of real leaves---cell walls (0.2--1~$\mu$m), chloroplasts (3--6~$\mu$m), intercellular air cavities (1--10~$\mu$m), vascular boundaries, and fungal hyphae during early infection---have characteristic dimensions comparable to VIS--NIR wavelengths. 
In this wavelength-scale regime, coherent electromagnetic effects play a significant role in shaping internal irradiance patterns and, consequently, the measured reflectance and transmittance.
Therefore, accurate modeling of light propagation in complex leaf tissues requires a full-wave treatment that directly solves Maxwell’s equations.\footnote{Full-wave optical modeling refers to numerical methods---such as finite-difference time-domain, finite element methods, or boundary integral formulations---that solve Maxwell’s equations without invoking high-frequency or ray-based approximations, thereby capturing interference, diffraction, guided modes, and resonant electromagnetic phenomena.}

\begin{figure}[htbp]
    \centering
    \subfloat[\centering Without a wedge]{
        \includegraphics[width=0.4\textwidth]{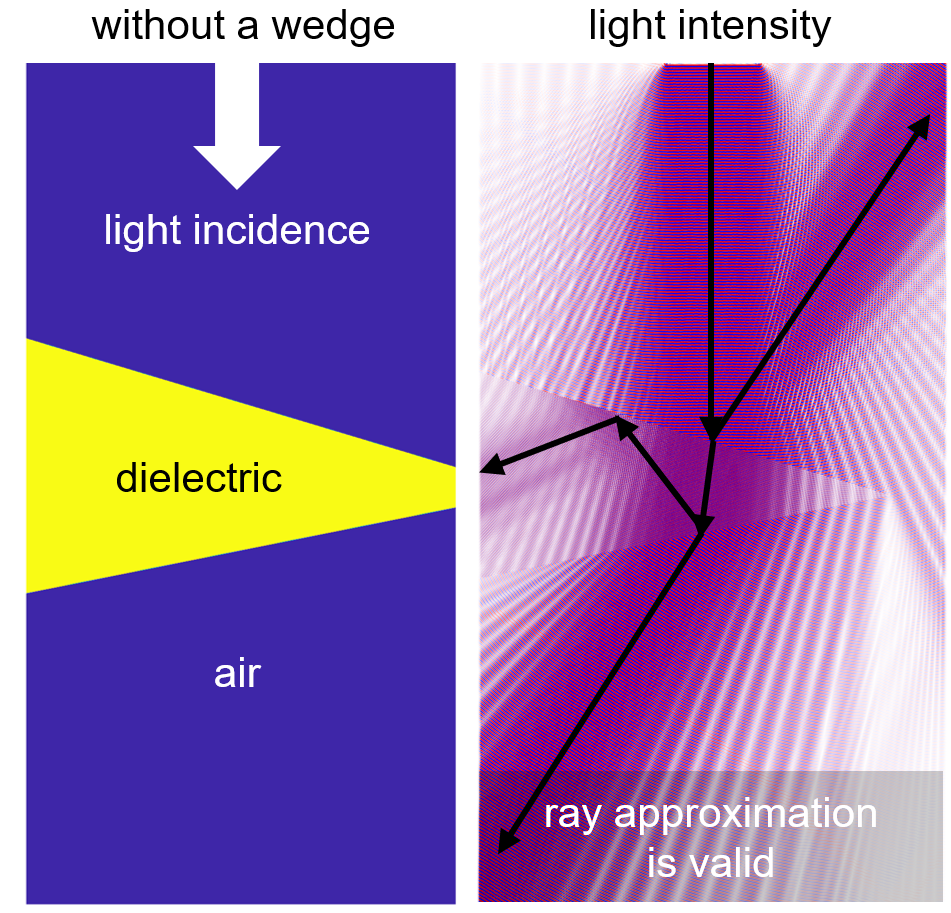}
    }
    \quad\quad
    \subfloat[\centering With a wedge]{
        \includegraphics[width=0.4\textwidth]{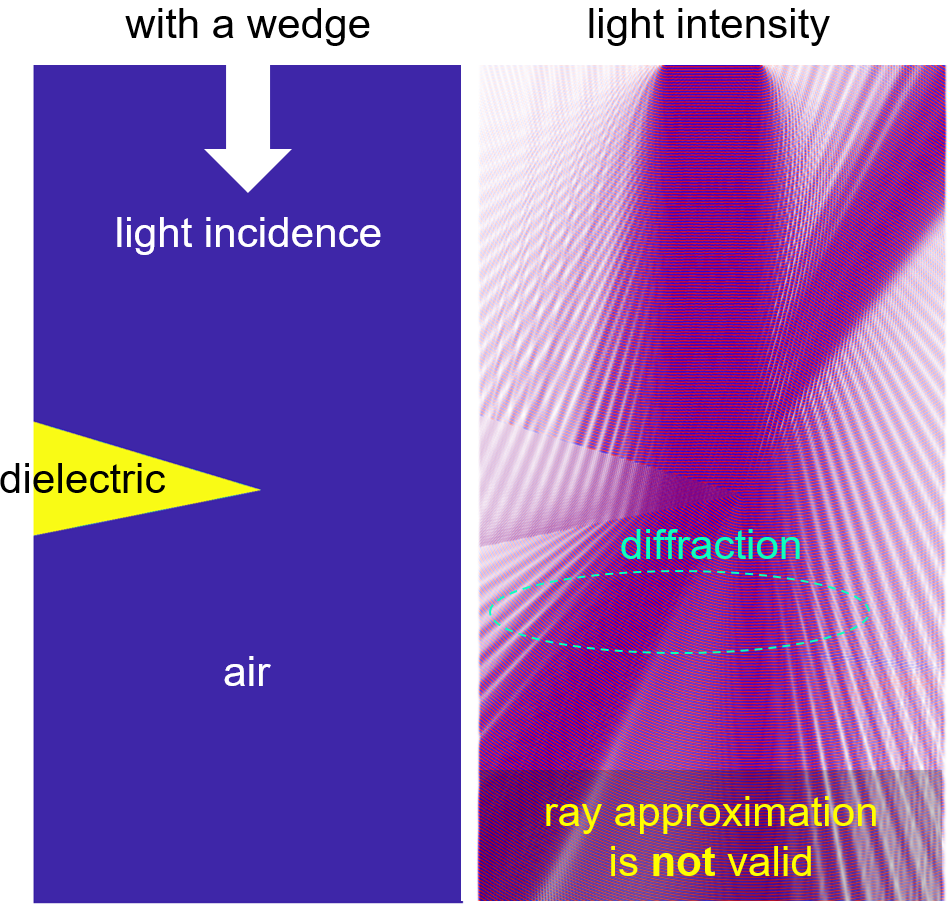}
    }
    \caption{Comparison of light scattering patterns (a) without and (b) with a wedge.  
    The wedge introduces strong diffraction and interference effects that cannot be described by ray-based approximations.}
    \label{fig:wedge}
\end{figure}

To illustrate this scale-dependent breakdown, consider the electrical length of a structure,
\[
\text{Electrical length} = \frac{D}{\lambda},
\]
where $D$ is the characteristic feature size. 
When $D/\lambda \gg 1$, scattering is weak and ray optics provides an accurate approximation. 
When $D/\lambda \sim 1$ or when sharp discontinuities are present, diffraction and interference dominate and ray models fail, as illustrated in Fig.~\ref{fig:wedge}. 
This situation closely mirrors the heterogeneous microstructures of plant leaves, motivating the use of full-wave numerical solvers such as Finite-Difference Time-Domain (FDTD).

\subsection{Contributions of this work}
To address these limitations, this work introduces a physics-grounded, full-wave optical modeling framework based on the FDTD method. 
Cross-sectional microscope images were digitized and segmented into anatomically faithful tissues---including the cuticle, epidermis, palisade and spongy mesophyll, vascular bundles, and stomata---and wavelength-dependent complex refractive indices were assigned using reported optical constants of water, pigments, and cell-wall materials.
Because the electrical length of a leaf spans several hundred wavelengths across the VIS--NIR spectrum, full-wave modeling requires substantial computational resources. 
To make such simulations tractable, we implemented an in-house CUDA-based FDTD solver and performed all simulations on graphics processing units (GPUs), achieving speedups of several tens compared to CPU-based execution.
GPU-accelerated simulations over 400--2,500~nm were carried out for healthy monocot and dicot leaves. 
The resulting reflectance and transmittance spectra closely matched PROSPECT-PRO predictions while revealing interference-induced modulations that lie beyond the capabilities of ray-based or radiative-transfer models, thereby underscoring the necessity of a wave-level EM treatment. 
To the best of our knowledge, no prior study has reported a full-wave optical model of leaf light scattering using anatomically realistic microstructures.
Finally, by modeling necrotrophic fungal infection through melanized hyphae penetrating the epidermis, we demonstrate that full-wave simulations can directly link microscale pathological alterations to macroscopic optical signatures, providing a physics-based foundation for hyperspectral detection of early-stage disease.

\section{Method}

\subsection{Problem description}

\begin{figure}[ht]
\centering
\includegraphics[width=0.75\linewidth]{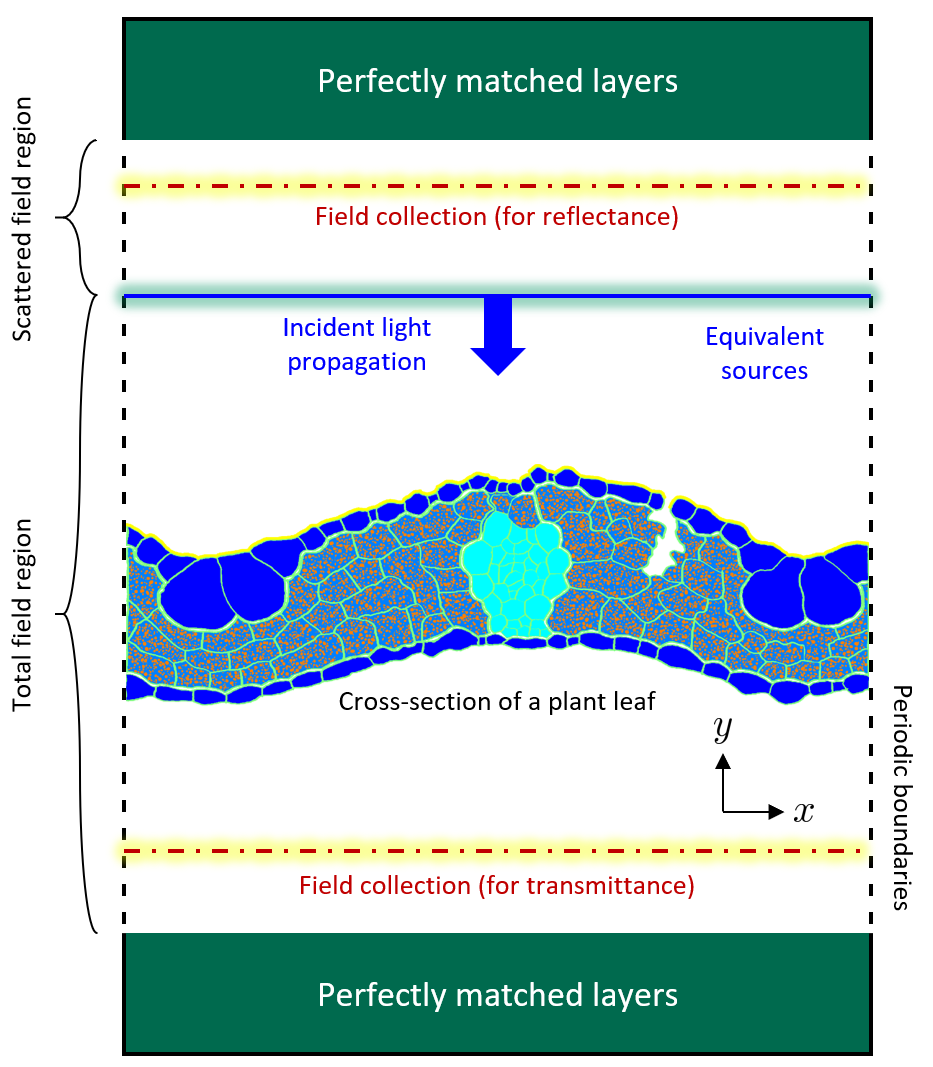}
\caption{Schematic of the FDTD simulation domain showing the cross-section of a plant leaf, total-field/scattered-field (TF/SF) regions, equivalent plane-wave excitation, field collection lines for reflectance and transmittance calculations, and boundary conditions (periodic boundaries along $x$ and perfectly matched layers along $y$).}
\label{fig:FDTD_sim_set}
\end{figure}

We develop a first-principles, full-wave optical modeling framework to compute the spectral reflectance and transmittance of plant leaves using FDTD simulations. For computational efficiency and as a proof-of-concept demonstration, the inherently three-dimensional leaf structure is approximated by a two-dimensional anatomical cross-section, as illustrated in Fig.~\ref{fig:FDTD_sim_set}. The internal anatomy of the leaf is embedded in the $xy$-plane, where $x$ denotes the horizontal axis and $y$ the vertical axis, and a monochromatic plane wave is incident along the $-y$ direction.

Because natural sunlight is broadband, thermal, and unpolarized, the incident field in the two-dimensional simulation is decomposed into two orthogonal components: P-polarization and S-polarization.  
Each polarization is simulated independently, and their results are later combined to represent the general case of unpolarized illumination.

\subsection{Maxwell's curl equations for S- and P-polarized illumination}

Plant leaves contain various types of cells---such as chloroplasts, vacuoles, and cell walls---whose optical responses are frequency dependent. 
These materials must therefore be modeled as dispersive media with complex refractive indices that vary with frequency.  
To incorporate such dispersion into FDTD simulations, one may employ the auxiliary differential equation formulation by introducing polarization-current terms~\cite{joseph1991direct,taflove2005computational,alsunaidi2009general}.  
The Lorentz--Drude--Sommerfeld model provides a physically rigorous description of dispersive behavior consistent with causality~\cite{chew2020lectures}, enabling broadband reflectance and transmittance to be computed from a single simulation through post-processing.

For the present proof-of-concept demonstration, however, we adopt a simpler approach:  
each wavelength is simulated independently using a monochromatic light excitation.  
Although this requires multiple FDTD runs to obtain a full spectrum, it greatly simplifies both formulation and implementation.  
For a dispersive medium with complex refractive index
\[
\tilde{n}(\omega)=n(\omega)+i\kappa(\omega),
\]
the corresponding relative permittivity and equivalent conductivity are
\begin{flalign}
\epsilon_r(\omega)
  &= \tilde{n}^2(\omega)
   = n^2(\omega) - \kappa^2(\omega) + i\,2 n(\omega)\kappa(\omega), \\[4pt]
\sigma(\omega)
  &= \omega\epsilon_0\,\Im\!\left[\epsilon_r(\omega)\right]
   = 2\omega\epsilon_0\,n(\omega)\kappa(\omega),
\end{flalign}
where $\epsilon_0$ is the permittivity of free space.

Using these material parameters, the phasor-domain Maxwell curl equations become
\begin{flalign}
\nabla \times \mathbf{E}(\mathbf{r},\omega)
  &= i\omega\mu_0\,\mathbf{H}(\mathbf{r},\omega)
     - \mathbf{M}(\mathbf{r},\omega), \\[4pt]
\nabla \times \mathbf{H}(\mathbf{r},\omega)
  &= \left[-i\omega\epsilon_r(\mathbf{r},\omega)\epsilon_0
     + \sigma(\mathbf{r},\omega)\right] \mathbf{E}(\mathbf{r},\omega)
     + \mathbf{J}(\mathbf{r},\omega),
\end{flalign}
where $\mathbf{M}$ and $\mathbf{J}$ denote magnetic- and electric-current excitations, respectively.  
All leaf tissues are assumed nonmagnetic and isotropic, though spatially inhomogeneous.

Transforming these equations into the time domain yields
\begin{flalign}
\nabla \times \mathbf{E}_\omega(\mathbf{r},t)
  &= -\mu_0\,\frac{\partial \mathbf{H}_\omega}{\partial t}
     - \mathbf{M}_\omega(\mathbf{r},t), \\[4pt]
\nabla \times \mathbf{H}_\omega(\mathbf{r},t)
  &= \epsilon_{r,\omega}(\mathbf{r})\epsilon_0\,
     \frac{\partial \mathbf{E}_\omega}{\partial t}
     + \sigma_\omega(\mathbf{r})\,\mathbf{E}_\omega(\mathbf{r},t)
     + \mathbf{J}_\omega(\mathbf{r},t),
\end{flalign}
where $\epsilon_{r,\omega}(\mathbf{r})=\epsilon_r(\mathbf{r},\omega)$ and 
$\sigma_\omega(\mathbf{r})=\sigma(\mathbf{r},\omega)$ are constant for a fixed excitation frequency.  
For brevity, we omit the subscript $\omega$ in the following.

Under these assumptions, the two-dimensional Maxwell curl equations for S- and P-polarized illumination can be written as follows.
For P-polarized case (or TE$_z$), $\mathbf{E}=\hat{\mathbf{x}}E_x + \hat{\mathbf{y}}E_y$ and $\mathbf{H}=\hat{\mathbf{z}}H_z$:
\begin{align}
\frac{\partial H_z}{\partial t}
  &= -\frac{1}{\mu_0}\left(
        \frac{\partial E_y}{\partial x}
        - \frac{\partial E_x}{\partial y}
     \right)
     - \frac{1}{\mu_0}M_z, \\[4pt]
\frac{\partial E_x}{\partial t}
  &= \frac{1}{\epsilon_0\epsilon_r}
     \frac{\partial H_z}{\partial y}
     - \frac{\sigma}{\epsilon_0\epsilon_r} E_x
     - \frac{1}{\epsilon_0\epsilon_r}J_x, \\[4pt]
\frac{\partial E_y}{\partial t}
  &= -\frac{1}{\epsilon_0\epsilon_r}
      \frac{\partial H_z}{\partial x}
     - \frac{\sigma}{\epsilon_0\epsilon_r} E_y
     - \frac{1}{\epsilon_0\epsilon_r}J_y.
\end{align}
For S-polarized case (or TM$_z$), $\mathbf{H}=\hat{\mathbf{x}}H_x + \hat{\mathbf{y}}H_y$ and $\mathbf{E}=\hat{\mathbf{z}}E_z$:
\begin{align}
\frac{\partial H_x}{\partial t}
  &= -\frac{1}{\mu_0}\frac{\partial E_z}{\partial y}
     - \frac{1}{\mu_0}M_x, \\[4pt]
\frac{\partial H_y}{\partial t}
  &=  \frac{1}{\mu_0}\frac{\partial E_z}{\partial x}
     - \frac{1}{\mu_0}M_y, \\[4pt]
\frac{\partial E_z}{\partial t}
  &=  \frac{1}{\epsilon_0\epsilon_r}
      \left(
        \frac{\partial H_y}{\partial x}
        - \frac{\partial H_x}{\partial y}
      \right)
     - \frac{\sigma}{\epsilon_0\epsilon_r}E_z
     - \frac{1}{\epsilon_0\epsilon_r}J_z.
\end{align}

\subsection{Two-dimensional finite-difference time-domain scheme}

\begin{figure}[htbp]
    \centering
    \subfloat[\centering TE\textsubscript{z} polarzation]{
        \includegraphics[width=0.45\textwidth]{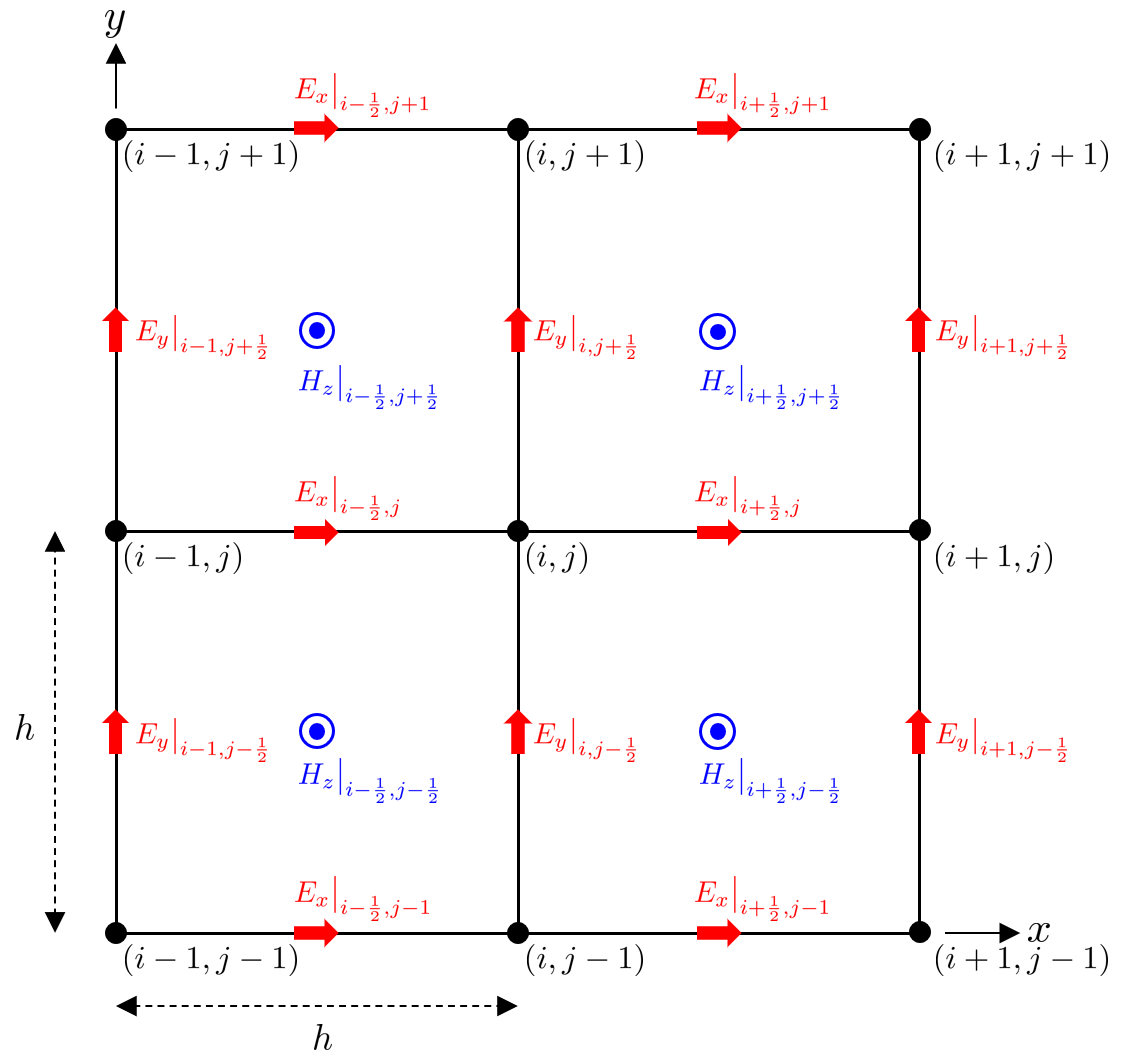}
        \label{fig:FDTD_discretization_sub1}
    }        
    \subfloat[\centering TM\textsubscript{z} polarzation]{
        \includegraphics[width=0.45\textwidth]{Figures/space_discretization_P_pol.png}
        \label{fig:FDTD_discretization_sub2}
    }    
    \caption{Schematic of the two-dimensional Yee grid illustrating the field component arrangement for (a) TE\textsubscript{z} and (b) TM\textsubscript{z} polarizations.}
    \label{fig:space_discretization}
\end{figure}

\begin{figure}[ht]
\centering
\includegraphics[width=0.45\linewidth]{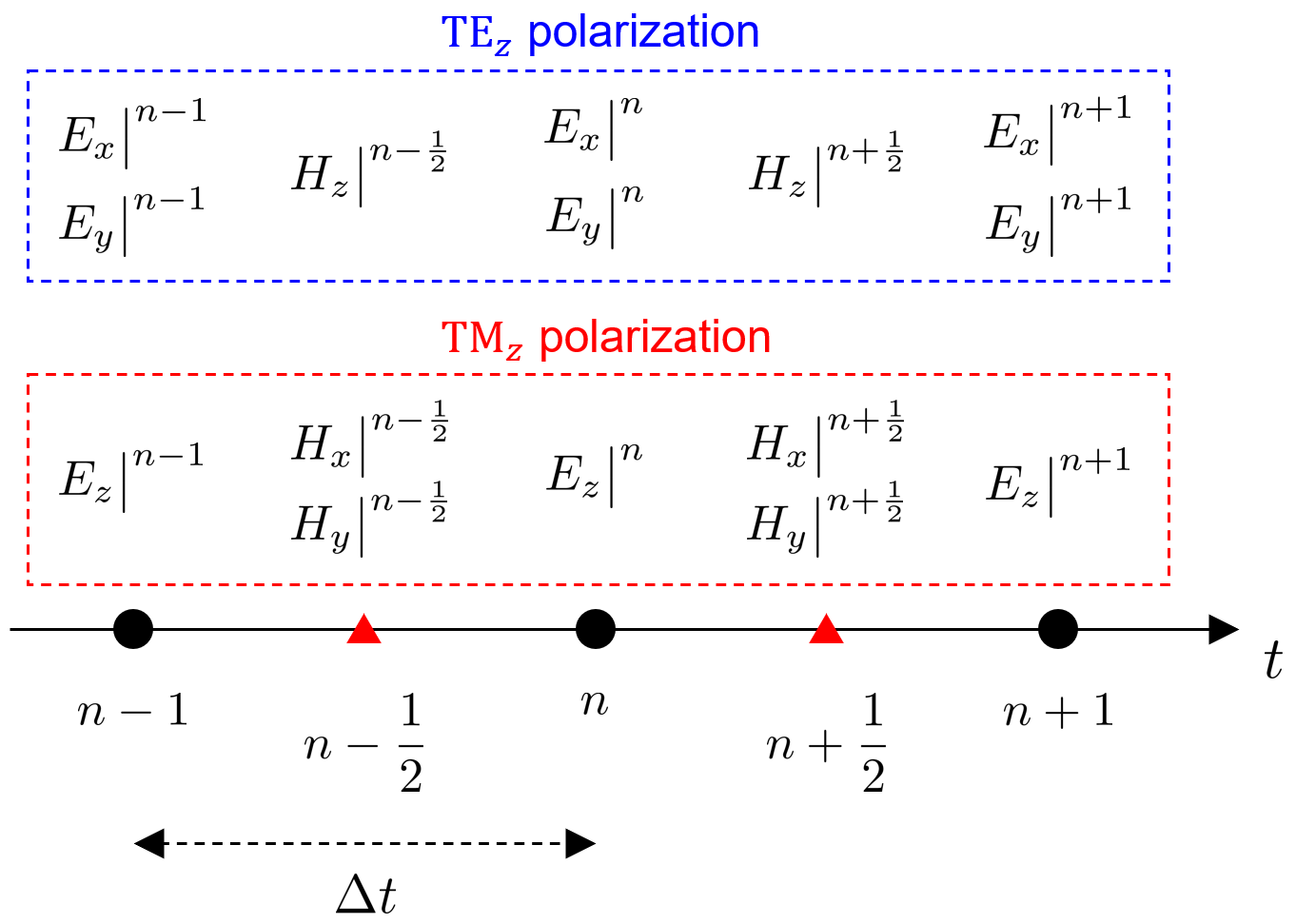}
\caption{Schematic illustration of the leapfrog time-stepping schemes for TE\textsubscript{z} and (b) TM\textsubscript{z} polarizations.}
\label{fig:time_discretization}
\end{figure}
    
The FDTD method numerically solves Maxwell’s curl equations by discretizing the EM fields on a staggered Yee grid and advancing them in time using a leap-frog integration scheme~\cite{yee1966numerical,taflove2005computational}. 
More specifically, spatial and temporal derivatives are approximated using finite differences, allowing the electric and magnetic field components to be updated sequentially in time in response to the prescribed electric and magnetic current densities.
Note also that the FDTD method updates all field quantities \textit{explicitly}, i.e., in a matrix-free manner. 
Moreover, the staggered-grid field components can be easily mapped into one-dimensional arrays, and each time-step involves only elementwise additions, subtractions, scalar multiplications, and Hadamard (elementwise) products. This structure makes the FDTD algorithm highly amenable to GPU acceleration.

The computational domain (or mesh) is discretized using Yee's lattice, and the grid points are defined as
\[
x_i \triangleq i\,h, \quad y_j \triangleq j\,h, \quad t^n \triangleq n\,\Delta t,
\]
where $h$ denotes the spatial grid spacing both in $x$ and $y$ directions, and $\Delta t$ represents the temporal step size. 
Subscripts and superscripts denote the spatial grid index and the time index, respectively, i.e.,
\begin{flalign}
A\bigr|_{i,j}^{n} \triangleq A(x_i, y_j, t^n).
\end{flalign}
For the TE$_z$ (or P) polarization, the discrete time-stepping relations are expressed as
\begin{align}
\bigl.H_z\bigr|^{\,n+\frac{1}{2}}_{i+\frac{1}{2},j+\frac{1}{2}} &=
\bigl.H_z\bigr|^{\,n-\frac{1}{2}}_{i+\frac{1}{2},j+\frac{1}{2}}
\nonumber \\
&-\frac{\Delta t}{\mu_0}
\left[
\frac{\bigl.E_y\bigr|^{\,n}_{i+1,j+\frac{1}{2}}-\bigl.E_y\bigr|^{\,n}_{i,j+\frac{1}{2}}}{h}
-\frac{\bigl.E_x\bigr|^{\,n}_{i+\frac{1}{2},j+1}-\bigl.E_x\bigr|^{\,n}_{i+\frac{1}{2},j}}{h}
+ M_z\bigr|_{i+\frac{1}{2},j+\frac{1}{2}}^{n}
\right], 
\\[6pt]
\bigl.E_x\bigr|^{\,n+1}_{i+\frac{1}{2},j} &=
C_e\bigr|_{i+\tfrac{1}{2},j}\,\bigl.E_x\bigr|^{\,n}_{i+\frac{1}{2},j}
+
C_h\bigr|_{i+\frac{1}{2},j}
\left[
\frac{
\bigl.H_z\bigr|^{\,n+\frac{1}{2}}_{i+\frac{1}{2},j+\frac{1}{2}}
-\bigl.H_z\bigr|^{\,n+\frac{1}{2}}_{i+\frac{1}{2},j-\frac{1}{2}}
}{h}
-
J_x\bigr|_{i+\frac{1}{2},j}^{n+\frac{1}{2}}
\right], \\[6pt]
\bigl.E_y\bigr|^{\,n+1}_{i,j+\frac{1}{2}} &=
C_e\bigr|_{i,j+\tfrac{1}{2}}\,\bigl.E_y\bigr|^{\,n}_{i,j+\frac{1}{2}}
-C_h\bigr|_{i,j+\tfrac{1}{2}}
\left[
\frac{
\bigl.H_z\bigr|^{\,n+\frac{1}{2}}_{i+\frac{1}{2},j+\frac{1}{2}}
-\bigl.H_z\bigr|^{\,n+\frac{1}{2}}_{i-\frac{1}{2},j+\frac{1}{2}}
}{h}
+
J_y\bigr|_{i,j+\frac{1}{2}}^{n+\frac{1}{2}}
\right],
\end{align}
where the conductivity-averaged coefficients for P-polarized fields are defined as
\begin{align}
C_e\bigr|_{i,j} &= 
\left(1 - \frac{\sigma\bigr|_{i,j}\Delta t}{2\epsilon_0\epsilon_r\bigr|_{i,j}}
\right)
\left(
1 + \frac{\sigma\bigr|_{i,j}\Delta t}{2\epsilon_0\epsilon_r\bigr|_{i,j}}\right)^{-1}, 
\qquad
C_h\bigr|_{i,j} = \left(1 + \frac{\sigma\bigr|_{i,j}\Delta t}{2\epsilon_0\epsilon_r\bigr|_{i,j}}\right)^{-1}.
\end{align}
These coefficients effectively average the conductivity term at times $t^n$ and $t^{n+1}$, ensuring numerical stability and consistent energy decay in lossy media.
For the TM$_z$ (or S) polarization, the discrete time-stepping relations are written as
\begin{align}
\bigl.E_z\bigr|^{\,n+\frac{1}{2}}_{i+\frac{1}{2},j+\frac{1}{2}} 
&=
C_e\bigr|_{i+\tfrac{1}{2},j+\tfrac{1}{2}}\,\bigl.E_z\bigr|^{\,n-\frac{1}{2}}_{i+\frac{1}{2},j+\frac{1}{2}}
\nonumber\\
&
\!\!\!\!\!\!\!\!\!\!
+
C_h\bigr|_{i+\tfrac{1}{2},j+\tfrac{1}{2}}
\left[
\frac{\bigl.H_y\bigr|^{\,n}_{i+1,j+\frac{1}{2}}-\bigl.H_y\bigr|^{\,n}_{i,j+\frac{1}{2}}}{h}
-\frac{\bigl.H_x\bigr|^{\,n}_{i+\frac{1}{2},j}-\bigl.H_x\bigr|^{\,n}_{i+\frac{1}{2},j-1}}{h}
- J_z\bigr|^{n}_{i+\frac{1}{2},j+\frac{1}{2}} 
\right], 
\\[6pt]
\bigl.H_x\bigr|^{\,n+1}_{i+\frac{1}{2},j} &=
\bigl.H_x\bigr|^{\,n}_{i+\frac{1}{2},j}
-\frac{\Delta t}{\mu_0}
\left[
\frac{
\bigl.E_z\bigr|^{\,n+\frac{1}{2}}_{i+\frac{1}{2},j+\frac{1}{2}}
-\bigl.E_z\bigr|^{\,n+\frac{1}{2}}_{i+\frac{1}{2},j-\frac{1}{2}}
}{h}
+
M_x\bigr|^{n+\frac{1}{2}}_{i+\frac{1}{2},j}
\right], \\[6pt]
\bigl.H_y\bigr|^{\,n+1}_{i,j+\frac{1}{2}} &=
\bigl.H_y\bigr|^{\,n}_{i,j+\frac{1}{2}}
+\frac{\Delta t}{\mu_0}
\left[
\frac{
\bigl.E_z\bigr|^{\,n+\frac{1}{2}}_{i+\frac{1}{2},j+\frac{1}{2}}
-\bigl.E_z\bigr|^{\,n+\frac{1}{2}}_{i-\frac{1}{2},j+\frac{1}{2}}
}{h}
-
M_y\bigr|^{n+\frac{1}{2}}_{i,j+\frac{1}{2}}
\right].
\end{align}
Again, the above scheme is matrix-free (or explicit) due to the use of the central-difference method, meaning that no matrix inversion is required.  
However, it is conditionally stable depending on the choice of the time step~$\Delta t$.  
The Courant--Friedrichs--Lewy (CFL) stability condition for the two-dimensional FDTD scheme \cite{courant1928partiellen,taflove2005computational} is given by
\[
\Delta t \le
\frac{1}{c\,\sqrt{\frac{1}{h^{2}} + \frac{1}{h^{2}}}},
\]
where $c$ denotes the speed of light.

\subsection{Boundary conditions of electromagnetic fields}
Since the electrical size of a plant leaf is large, the computational
domain must be truncated while ensuring physically consistent boundary
conditions. 
To emulate lateral periodicity of the leaf cross-section,
periodic boundary conditions (PBCs) \cite{harms1994implementation,taflove2005computational}
are applied along the $x$-direction.

For the TE$_z$ polarization, both $E_x$ and $E_y$ exist; however, the
tangential electric field on the lateral boundary is $E_y$.
Thus, the periodicity condition is imposed as
\[
E_y(x+L_x, y, t) = E_y(x, y, t) 
\]
where $L_x$ denotes the horizontal extent of the domain. 
In discretized form,
\begin{flalign}
\bigl.E_y\bigr|^{\,n}_{N_{g,x},\,j+\frac{1}{2}}
=
\bigl.E_y\bigr|^{\,n}_{0,\,j+\frac{1}{2}}
\end{flalign}
for $j = 0, 1, \dots, N_{g,y}-1$.
Similarly, for the TM$_z$ polarization, the tangential magnetic field on
the periodic boundary is $H_y$, so
\[
H_y(x+L_x, y, t) = H_y(x, y, t)
\]
which becomes
\begin{flalign}
\bigl.H_y\bigr|^{\,n}_{N_{g,x},\,j+\frac{1}{2}}
=
\bigl.H_y\bigr|^{\,n}_{0,\,j+\frac{1}{2}}
\end{flalign}
for $j = 0, 1, \dots, N_{g,y}-1$.

Because of these constraints, the FDTD update equations at the left and right boundaries must wrap field values across the domain. 
For TE$_z$, the quantity $\bigl.H_z\bigr|^{n+\frac12}_{-\frac12,j+\frac12}$ is replaced by its periodic counterpart
$\bigl.H_z\bigr|^{n+\frac12}_{N_{g,x}-\frac12,j+\frac12}$, yielding the
left--boundary update:
\[
\bigl.E_y\bigr|^{n+1}_{0,\,j+\frac12}
=
C_e\,\bigl.E_y\bigr|^{n}_{0,\,j+\frac12}
-
C_h
\left[
\frac{
\bigl.H_z\bigr|^{n+\frac12}_{\frac12,\,j+\frac12}
-
\bigl.H_z\bigr|^{n+\frac12}_{N_{g,x}-\frac12,\,j+\frac12}
}{h}
\right].
\]
An analogous wrapping is applied at the right boundary. Since these two
boundary update equations are identical up to index wrapping, they can be
expressed compactly as
\[
\bigl.E_y\bigr|^{\,n+1}_{\text{lateral},\,j+\frac12}
=
C_e\,\bigl.E_y\bigr|^{\,n}_{\text{lateral},\,j+\frac12}
-
C_h
\left[
\frac{
\bigl.H_z\bigr|^{n+\frac12}_{\frac12,\,j+\frac12}
-
\bigl.H_z\bigr|^{n+\frac12}_{N_{g,x}-\frac12,\,j+\frac12}
}{h}
\right].
\]
The same procedure applies to the TM$_z$ polarization:
\[
\bigl.H_y\bigr|^{\,n+1}_{\text{lateral},\,j+\frac12}
=
\bigl.H_y\bigr|^{\,n}_{\text{lateral},\,j+\frac12}
+
\frac{\Delta t}{\mu_0}
\left[
\frac{
\bigl.E_z\bigr|^{n+\frac12}_{\frac12,\,j+\frac12}
-
\bigl.E_z\bigr|^{n+\frac12}_{N_{g,x}-\frac12,\,j+\frac12}
}{h}
\right].
\]
All other update equations referencing boundary nodes follow the same
principle of index wrapping.

At the top and bottom of the computational domain, perfectly matched layers (PMLs) \cite{berenger1994perfectly,gedney1996anisotropic,chew19943d,teixeira1999differential,taflove2005computational} are employed to absorb outgoing waves and suppress spurious reflections.
Various formulations exist for implementing PMLs, including the stretched-coordinate PML, the uniaxial PML, the diagonally anisotropic PML, and the convolutional PML.
In this work, we adopt the split-field PML \cite{berenger1994perfectly}, which offers excellent absorption over a wide angular and frequency range while ensuring numerical stability and seamless integration into the standard FDTD update procedure.

\subsection{Excitation of incident plane wave at normal incidence}

To inject a normally incident plane wave without introducing spurious reflections, we employ the total-field/scattered-field (TF/SF) formulation \cite{merewether2007implementing,taflove2005computational}.  
The domain is divided into a total-field (TF) region, where both the incident and scattered fields exist, and a surrounding scattered-field (SF) region, where only scattered fields propagate.  
The TF/SF interface acts as a fictitious boundary on which equivalent electric and magnetic current sources are enforced so that the TF region contains the prescribed incident wave while the SF region contains no incident fields.

For TE\textsubscript{z} polarization, the incident fields are
\begin{align}
\mathbf{E}_{\text{inc}}^{\text{TE}}(x,y,t) &= 
\hat{\mathbf{x}}\,\cos(\omega t - k y), \\
\mathbf{H}_{\text{inc}}^{\text{TE}}(x,y,t) &= 
\hat{\mathbf{z}}\,\frac{1}{\eta_0}\cos(\omega t - k y),
\end{align}
where the wave propagates in the $-y$ direction.  
Let $\hat{\mathbf{n}}=-\hat{\mathbf{y}}$ be the outward normal from the TF region into the SF region.  
The equivalent electric and magnetic current sources imposed on the TF/SF boundary are
\begin{align}
\mathbf{J}_{\text{eqv}}^{\text{TE}} &= \hat{\mathbf{n}} \times 
\mathbf{H}_{\text{inc}}^{\text{TE}}, \\
\mathbf{M}_{\text{eqv}}^{\text{TE}} &= -\,\hat{\mathbf{n}} \times 
\mathbf{E}_{\text{inc}}^{\text{TE}}.
\end{align}
These currents ensure that the incident field exists only within the TF region, thereby enabling accurate computation of reflection and transmission.

For TM\textsubscript{z} polarization, the incident plane wave is instead given by
\begin{align}
\mathbf{E}_{\text{inc}}^{\text{TM}}(x,y,t) &=
\hat{\mathbf{z}}\cos(\omega t - k y), \\
\mathbf{H}_{\text{inc}}^{\text{TM}}(x,y,t) &=
-\hat{\mathbf{x}}\,\frac{1}{\eta_0}\cos(\omega t - k y),
\end{align}
and substituting these fields into the same TF/SF equivalent-current expressions yields the corresponding TM-equivalent sources.

\subsection{Reflectance and Transmittance Calculation}

After the fields reach steady state, the reflected and transmitted signals are sampled along two horizontal lines located at 
$y = y_r$ and $y = y_t$, respectively, over multiple time periods.  
For TE\textsubscript{z} polarization, applying the discrete Fourier transform (DFT) to the recorded time-domain fields yields
\begin{align}
\tilde{E}_{\mathrm{ref}}(x,\omega) &= \mathcal{F}\{E_x(x,y_r,t)\}, \\
\tilde{E}_{\mathrm{trs}}(x,\omega) &= \mathcal{F}\{E_x(x,y_t,t)\}.
\end{align}
Because the simulation enforces periodicity in the $x$-direction, the spectral fields admit a Bloch expansion \cite{xie2006transmission}:
\begin{align}
\tilde{E}_{\mathrm{ref}}(x,\omega)
&= \sum_{m=-\infty}^{\infty}
E_{\mathrm{ref},m}(\omega)\,e^{i k_{x,m} x}, \\
\tilde{E}_{\mathrm{trs}}(x,\omega)
&= \sum_{m=-\infty}^{\infty}
E_{\mathrm{trs},m}(\omega)\,e^{i k_{x,m} x},
\end{align}
where the transverse Bloch wavenumber is 
\[
k_{x,m} = \frac{2\pi m}{L_x}.
\]
A Bloch order $m$ corresponds to a propagating mode only when its longitudinal wavenumber
\[
k_{y,m} = \sqrt{k_0^2 - k_{x,m}^2}
\]
is real.  
Let $M_{\max}$ denote the largest integer such that $k_{y,m}$ remains real.  
Then, summing all propagating orders yields the reflectance and transmittance:
\begin{align}
R(\omega)
&=
\frac{
\displaystyle\sum_{m=-M_{\max}}^{M_{\max}}
\left|E_{\mathrm{ref},m}(\omega)\right|^2
}{
\left|E_{\mathrm{inc},0}(\omega)\right|^2
},\\[4pt]
T(\omega)
&=
\frac{
\displaystyle\sum_{m=-M_{\max}}^{M_{\max}}
\left|E_{\mathrm{trs},m}(\omega)\right|^2
}{
\left|E_{\mathrm{inc},0}(\omega)\right|^2
}.
\end{align}
Here, $E_{\mathrm{inc},0}(\omega)$ is the spectral amplitude of the incident zeroth-order Bloch mode.
The absorption spectrum is then obtained from energy conservation:
\[
A(\omega) = 1 - R(\omega) - T(\omega).
\]
For TM\textsubscript{z} polarization, the same procedure applies by replacing  
$E_x(x,y_{r/t},t)$ with $E_z(x,y_{r/t},t)$ in the above formulation.

\subsection{Reconstruction of internal plant cross-section structure}
To accurately simulate light scattering within plant leaves, the internal microstructure must be represented with high anatomical fidelity. 
To this end, we reconstructed the leaf geometry by segmenting microscope cross-sectional images into their constituent anatomical regions. 
Representative cross-sectional images of dicot and monocot leaves were obtained from \cite{moffitt_leaf_structure} and \cite{da2020increased}, respectively. 
Using a SAMSUNG Galaxy Tab S7, a human rater manually delineated the boundaries of each anatomical layer with distinct colors. 
Following the manual annotation, digital image-processing techniques were applied to extract the boundary curves and convert them into closed polygonal regions suitable for numerical meshing and subsequent FDTD simulation.

During the manual annotation process, anatomical components such as the \textit{cuticle}, 
\textit{cell wall}, \textit{epidermal cells}, \textit{palisade mesophyll}, 
\textit{spongy mesophyll}, \textit{vascular bundles}, \textit{veins}, 
\textit{bulliform cells}, \textit{stroma}, and \textit{lower epidermal cells} 
were delineated. 
Due to the limited resolution of the microscope images, individual \textit{chloroplasts} 
could not be identified or manually traced. 
To compensate for this limitation, chloroplasts were modeled as randomly inserted 
\textit{elliptical shapes} of varying sizes within the \textit{mesophyll} and 
\textit{stroma} regions. 
The assumed chloroplast concentration was reflected in both the number and size 
distribution of the inserted ellipses.

Each chloroplast was represented as a rotated ellipse, corresponding to the 
projected cross-section of an oblate spheroid embedded within the mesophyll tissue. 
The major and minor axes are denoted by $a$ and $b$, respectively, and the rotation 
angle by~$\theta$. 
To capture natural morphological variability, the parameters 
$a$, $b$, and $\theta$ were treated as random variables. 
The orientation angle~$\theta$ followed a uniform distribution on~$[0, 2\pi)$.
The major axis was sampled as
\begin{equation}
    a_{\mathrm{mean}} = \mu_a + \sigma_a (2 r_1 - 1),
\end{equation}
where $\mu_a = 0.8~\mu\mathrm{m}$ is the mean chloroplast radius, 
$\sigma_a = 0.2~\mu\mathrm{m}$ the standard deviation, 
and $r_1 \sim U(0,1)$ is a uniformly distributed random number. 
The minor axis was then perturbed relative to $a_{\mathrm{mean}}$:
\begin{equation}
    b_{\mathrm{mean}} = a_{\mathrm{mean}} \bigl[1 + (r_2 - 0.5)\bigr],
\end{equation}
with $r_2 \sim U(0,1)$. 
This stochastic parameterization introduces sample-to-sample geometric variability 
while maintaining statistical consistency in chloroplast size distributions.
The chloroplast ellipses were randomly positioned within the mesophyll region, 
from the outer boundary to a depth $l_{\mathrm{chl}}$, using random centroids and 
samples of $(a, b, \theta)$. 
For healthy dicot leaves, the chloroplast filling fraction was controlled such that 
the total chloroplast area $A_{\mathrm{chl}}$ occupied approximately 20\% of the 
total mesophyll area $A_{\mathrm{mes}}$. 
Pigment concentrations were set to 
$C_{\mathrm{ab}} = 26{,}910~\mu\mathrm{g\,cm^{-3}}$ 
($\approx 30~\mathrm{mM}$) for chlorophyll~a+b and 
$C_{\mathrm{car}} = 3{,}221.4~\mu\mathrm{g\,cm^{-3}}$ 
($\approx 6~\mathrm{mM}$) for carotenoids.

\begin{figure}[htbp]
    \centering
    \subfloat[\centering Dicot]{
        \includegraphics[width=0.75\textwidth]{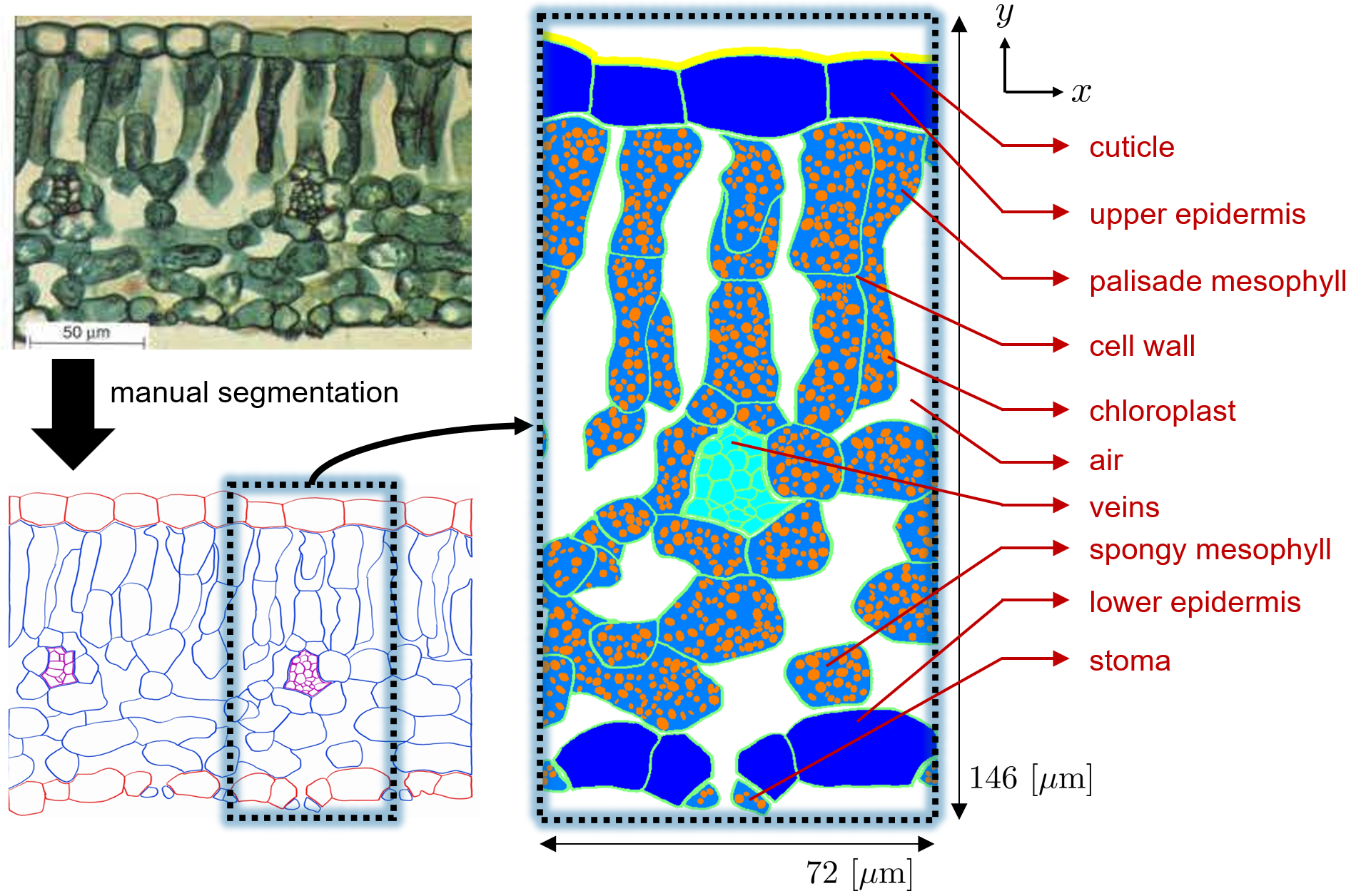}
        \label{fig:manual_segmentation_sub1}
    }    
    \\
    \subfloat[\centering Monocot]{
        \includegraphics[width=1\textwidth]{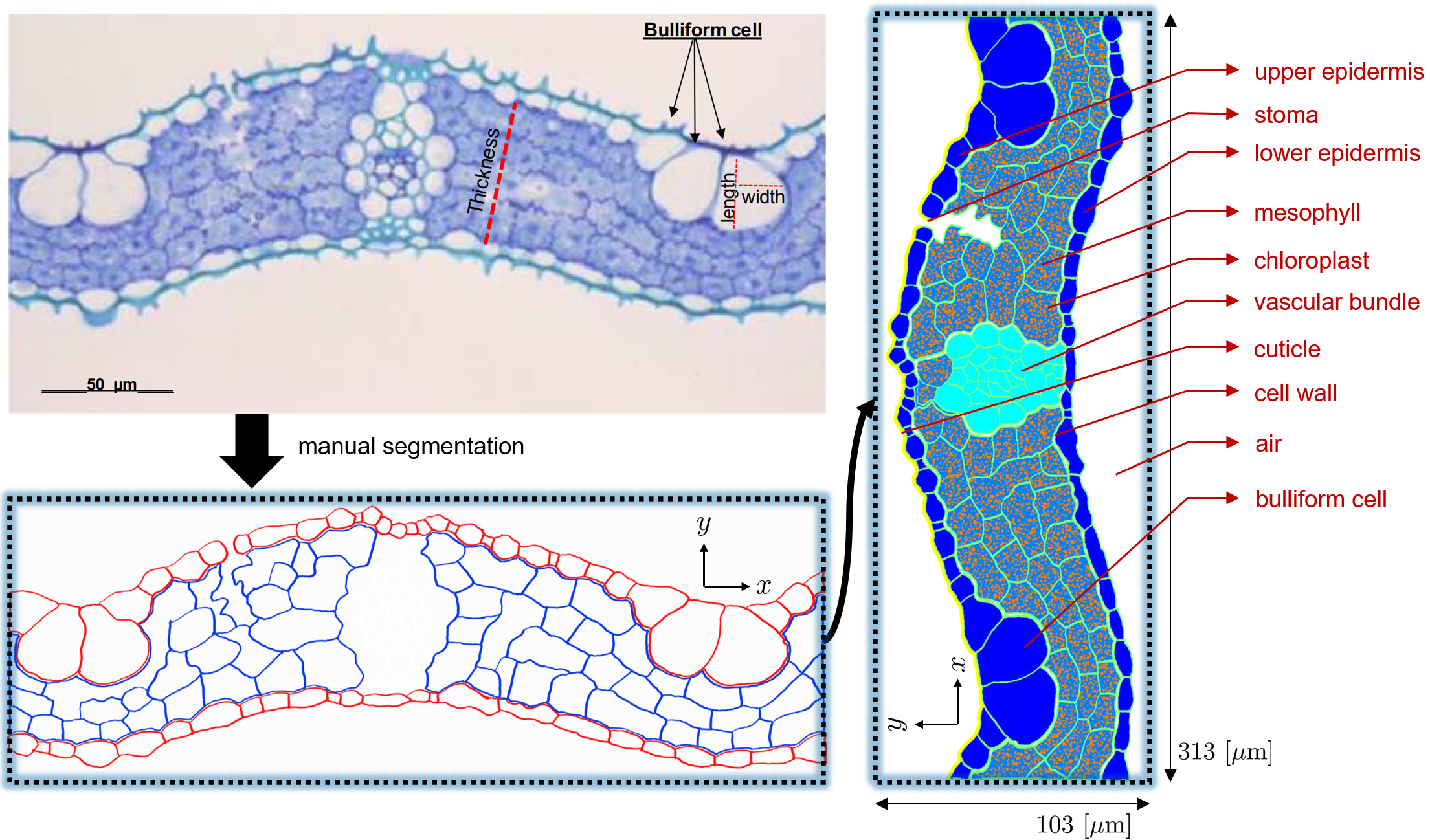}
        \label{fig:manual_segmentation_sub2}
    }
    \caption{Procedure for reconstructing the internal geometry from microscope images of (a) dicot and (b) monocot leaf samples.}
    \label{fig:manual_segmentation}
\end{figure}

The overall procedure used to reconstruct the internal plant structure for both dicot and monocot samples is illustrated in Fig.~\ref{fig:manual_segmentation}.

\subsection{Optical Properties of Plant Tissues}
The optical properties of materials are described by their complex refractive indices. 
When light propagates through media with different complex refractive indices, its reflection, transmission, and scattering characteristics vary significantly depending on both the refractive indices and the geometrical structure of the interfaces. 
Therefore, to accurately understand the light-scattering phenomena within the internal structure of a plant leaf, 
it is essential to model the complex refractive indices of each anatomical component correctly.

Complex refractive indices can be classified according to their physical characteristics \cite{chew2020lectures}, such as: (1) dispersion: whether the refractive index varies with frequency, (2) isotropy: whether the refractive index is the same in all directions, (3) inhomogeneity: whether the refractive index varies spatially, and (4) linearity: whether the dielectric polarization varies linearly with the incident electric field.
In this proof-of-concept study, to focus on the fundamental scattering mechanisms, all anatomical components of the leaf are assumed to be isotropic and linear.

In terms of optical behavior, plant tissues are primarily composed of water, pigments (chlorophylls and carotenoids), and structural cell-wall materials \cite{jacquemoud2019leaf,jacquemoud1996estimating,feret2008prospect,feret2017prospect,jacquemoud1990prospect,pfundel2008optical,vogelman1996leaves,gausman1973optical,gausman1974refractive,vogelmann1994light,xiao2016influence,roth2023plant,yang2023shifting}.
Accordingly, we classify four representative materials with distinct complex refractive indices to define the optical properties of each anatomical region in the leaf: (1) cuticle, (2) cell wall, (3) pigments, and (4) water.

The \textit{cuticle} is the outermost layer of the leaf surface, 
primarily composed of cutin and waxy hydrocarbons that serve as a protective barrier against water loss.  
Following~\cite{vogelmann2014leaf,peters2023characterization,gould1996physical,chen2012kramers}, the refractive index of the cuticle is modeled as
\[
n = 1.45.
\]

The \textit{cell wall} surrounds each plant cell and provides mechanical strength and rigidity. 
In potato and soybean leaves, it mainly consists of cellulose microfibrils, hemicellulose, pectin, and water.  
Based on reported measurements~\cite{gausman1974refractive,kumar1973reflectance,kumar1973light,green1958structural,woolley1971reflectance,nicoletti2000optical}, 
the refractive index of the cell wall is modeled as approximately
\[
n = 1.52.
\]

The complex refractive index of water varies with wavelength. 
In this study, both the real and imaginary parts of the water refractive index, 
denoted as $n_{\mathrm{wat}}(\lambda)$ and $\kappa_{\mathrm{wat}}(\lambda)$, 
were adopted from the PROSPECT-PRO dataset~\cite{feret2021prospect}.

\begin{figure}[htbp]
    \centering
    \subfloat[\centering specific absorption coefficient of Chlorophyll a+b and Carotenoids]{
        \includegraphics[width=0.32\textwidth]{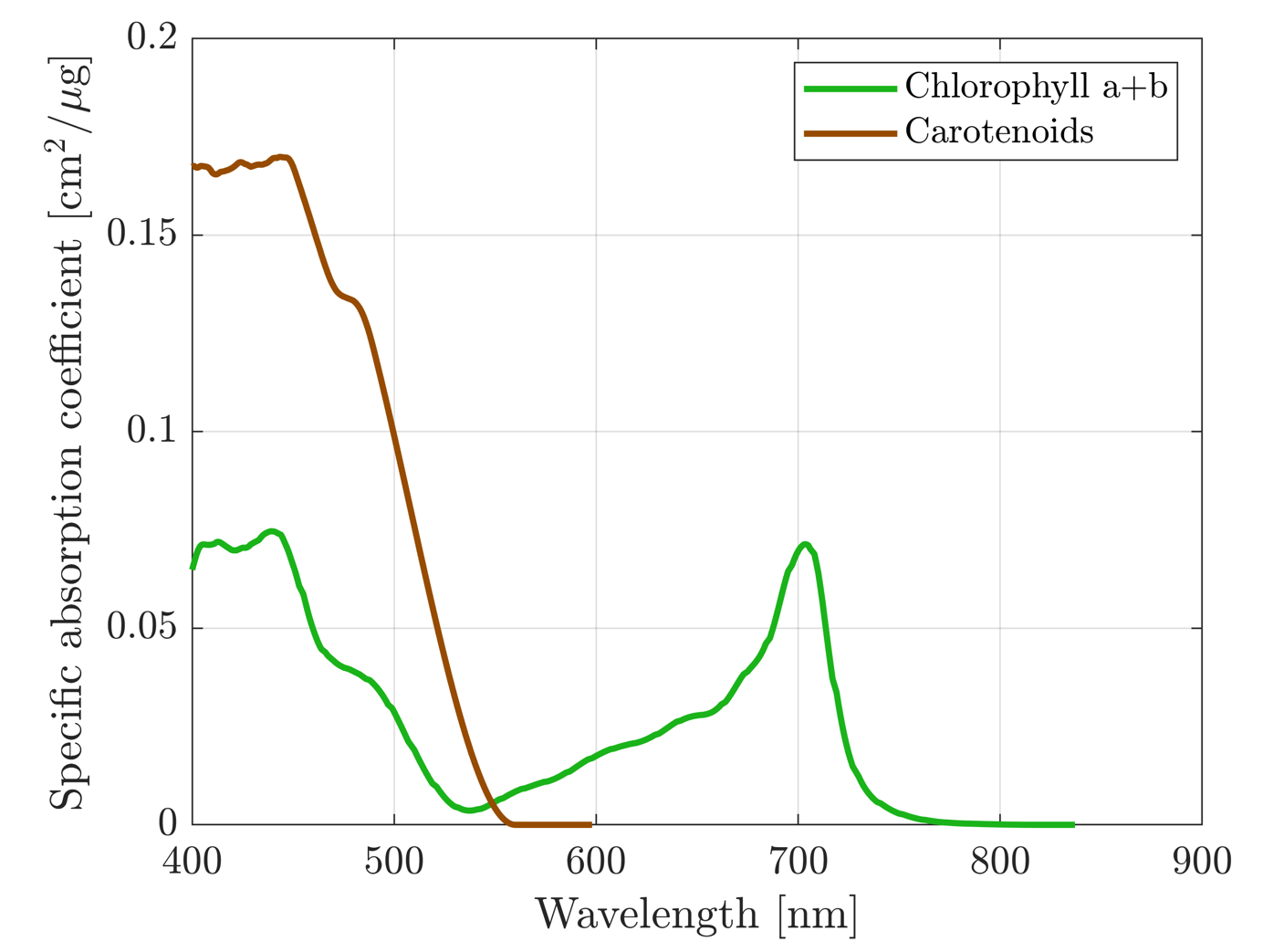}
        \label{fig:material_property_sub1}
    }        
    \subfloat[\centering extinction coefficient of pigments and water]{
        \includegraphics[width=0.32\textwidth]{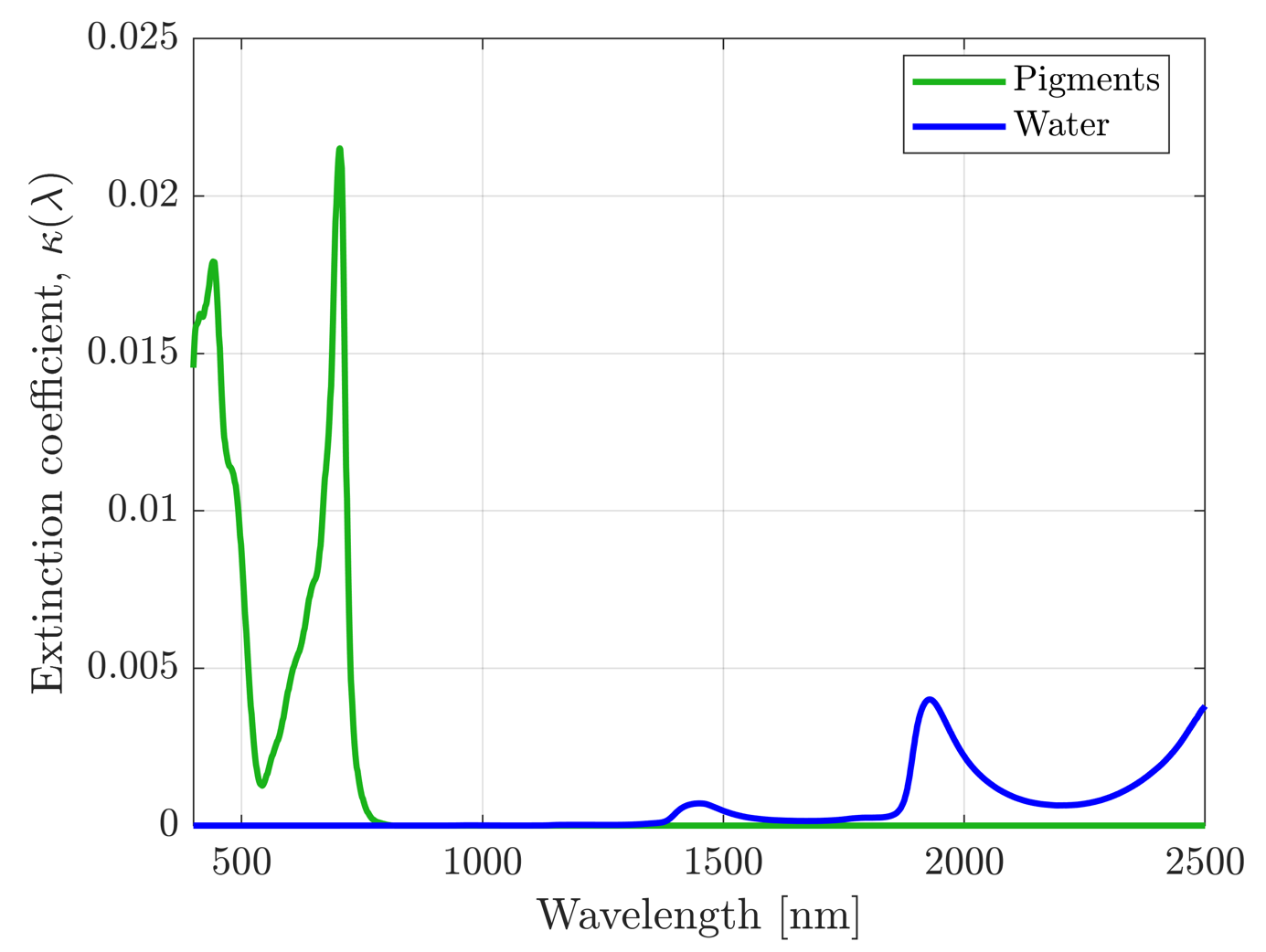}
        \label{fig:material_property_sub2}
    }
    \subfloat[\centering refractive index of cell wall, cuticle, pigments, and water]{
        \includegraphics[width=0.32\textwidth]{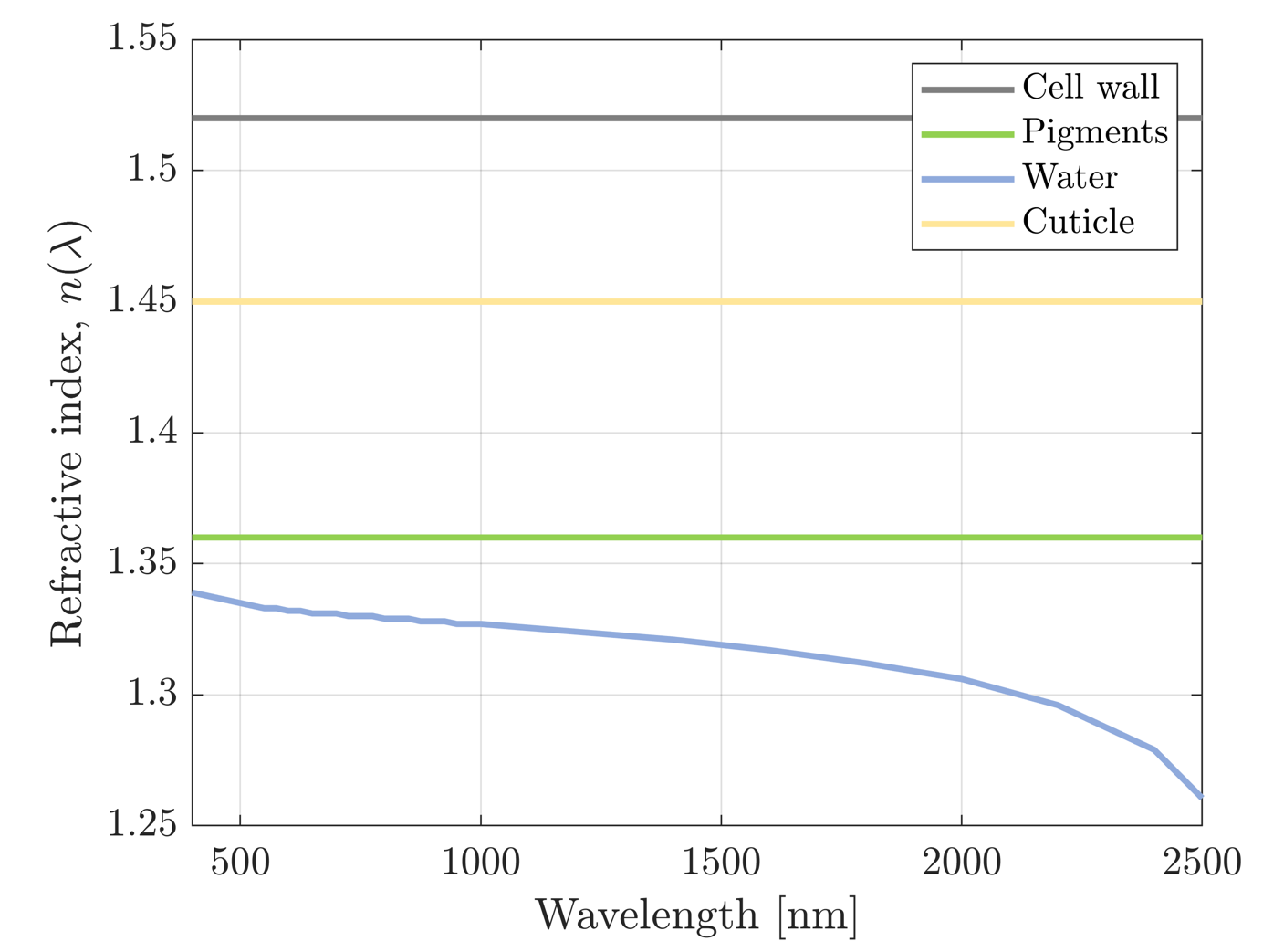}
        \label{fig:material_property_sub3}
    }
    \caption{Modeling of the optical properties of key materials composing the tissues of a plant leaf: (a) specific absorption coefficients of chlorophyll \textit{a+b} and carotenoids, (b) extinction coefficients of pigments (considering their concentrations) and water, and (c) refractive indices of the cell wall, cuticle, pigments, and water.}
    \label{fig:material_property}
\end{figure}

\textit{Pigments} are color-bearing molecules distributed primarily in the mesophyll region and play a central role in photosynthesis.  
In healthy plant leaves, the dominant pigments are \textit{chlorophyll~a}, \textit{chlorophyll~b}, 
and \textit{carotenoids} \cite{feret2021prospect}. 
These pigments exhibit distinct absorption spectra in the visible range, which determine the characteristic absorptance, reflectance, and transmittance of plant leaves.
To quantify their absorption, the \textit{specific absorption coefficient} (SAC) 
[cm\textsuperscript{2}\,µg\textsuperscript{-1}] 
is commonly employed by accounting for the pigment concentration within chloroplast regions. 
The SAC spectra of chlorophyll~a+b and carotenoids are shown in Fig.~\ref{fig:material_property}.
The complex refractive indices of each pigment were modeled as \cite{feret2021prospect,krekov2009radiative,datt1998remote}
\[
n_{\mathrm{chl}} = 1.36, \qquad 
\kappa_{\mathrm{chl\_ab}}(\lambda) = \frac{\mathrm{SAC}_{\mathrm{chl\_ab}}(\lambda)\,C_{\mathrm{ab}}}{0.01}
\frac{\lambda}{4\pi},
\]
\[
n_{\mathrm{car}} = 1.36, \qquad 
\kappa_{\mathrm{car}}(\lambda) = \frac{\mathrm{SAC}_{\mathrm{car}}(\lambda)\,C_{\mathrm{car}}}{0.01}
\frac{\lambda}{4\pi}.
\]
Finally, the overall complex refractive index of the pigment region was defined as
\[
n_{\mathrm{pig}}(\lambda) = 1.36, \qquad
\kappa_{\mathrm{pig}}(\lambda) = 
\kappa_{\mathrm{chl\_ab}}(\lambda) +
\kappa_{\mathrm{car}}(\lambda) +
\kappa_{\mathrm{wat}}(\lambda).
\]

The resulting wavelength-dependent complex refractive indices of the cuticle, cell wall, pigment, and water layers are summarized in Fig.~\ref{fig:material_property}.

\subsection{Implementation with GPU Acceleration and Computing Resources}

\begin{figure}[ht]
\centering
\includegraphics[width=0.75\linewidth]{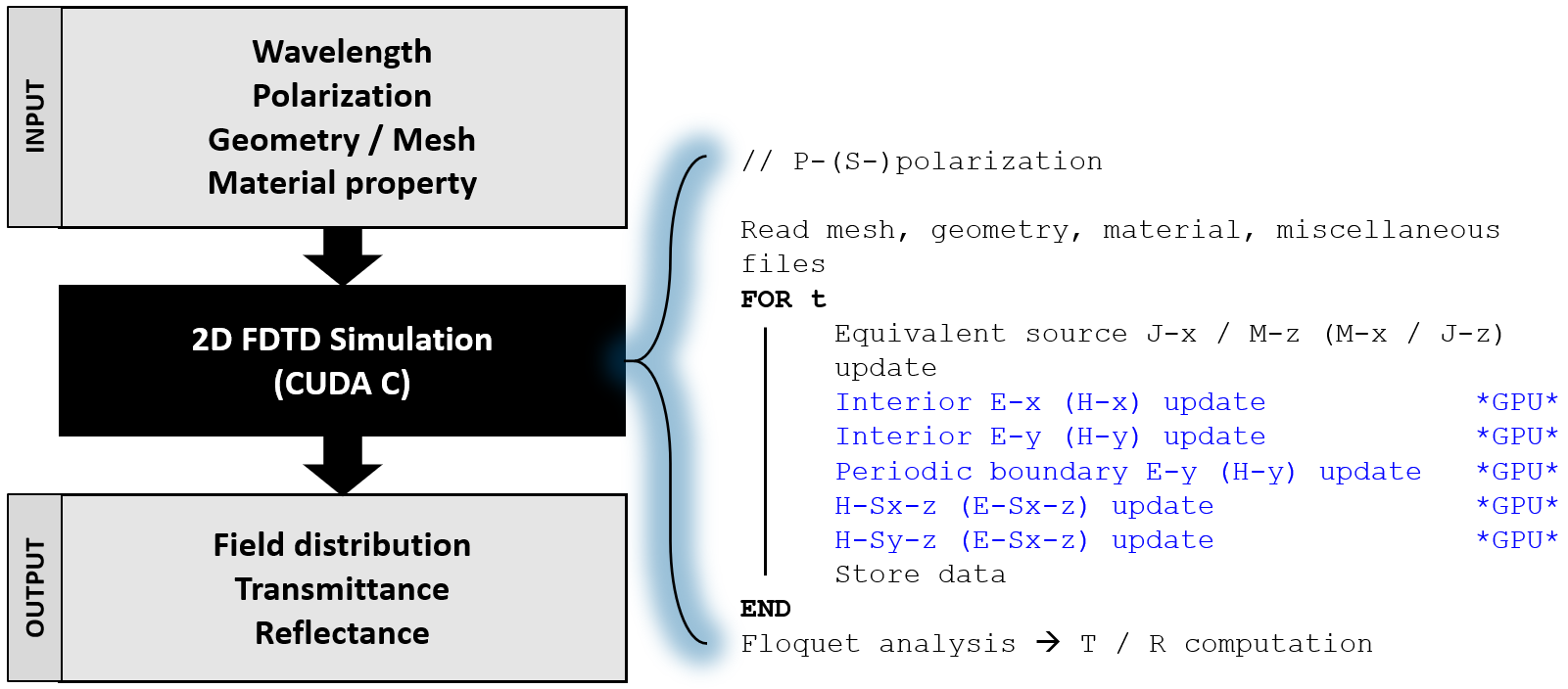}
\caption{In-house FDTD code implementation with GPU acceleration: FLARE-X.}
\label{fig:flarex}
\end{figure}

To enhance computational performance, we developed an in-house GPU-parallelized FDTD solver in CUDA~C and executed it in a Linux environment (Ubuntu on Windows~11 WSL).
A single GPU was used to perform high-throughput time-domain updates, as illustrated in Fig.~\ref{fig:flarex}.
The developed solver, termed \textit{FLARE-X} (FDTD Leaf-light Analysis with Rapid Execution on GPU Acceleration), is optimized for efficient simulation of light-scattering phenomena within anatomically realistic planar leaf microstructures.

All simulations were conducted on a workstation equipped with dual Intel\textsuperscript{\textregistered} Xeon\textsuperscript{\textregistered} Gold~6430 processors (2.10~GHz, 32~cores each) and 1.0~TB of DDR5 system memory.
GPU acceleration was provided by a single NVIDIA RTX~A6000 Ada graphics card with 48~GB of VRAM.
The operating environment was 64-bit Ubuntu running under Windows~11 WSL (Windows Subsystem for Linux).
This configuration enabled large-scale FDTD simulations to be executed efficiently within practical runtimes.

\section{Validation}
\begin{figure}[ht]
\centering
\includegraphics[width=0.5\linewidth]{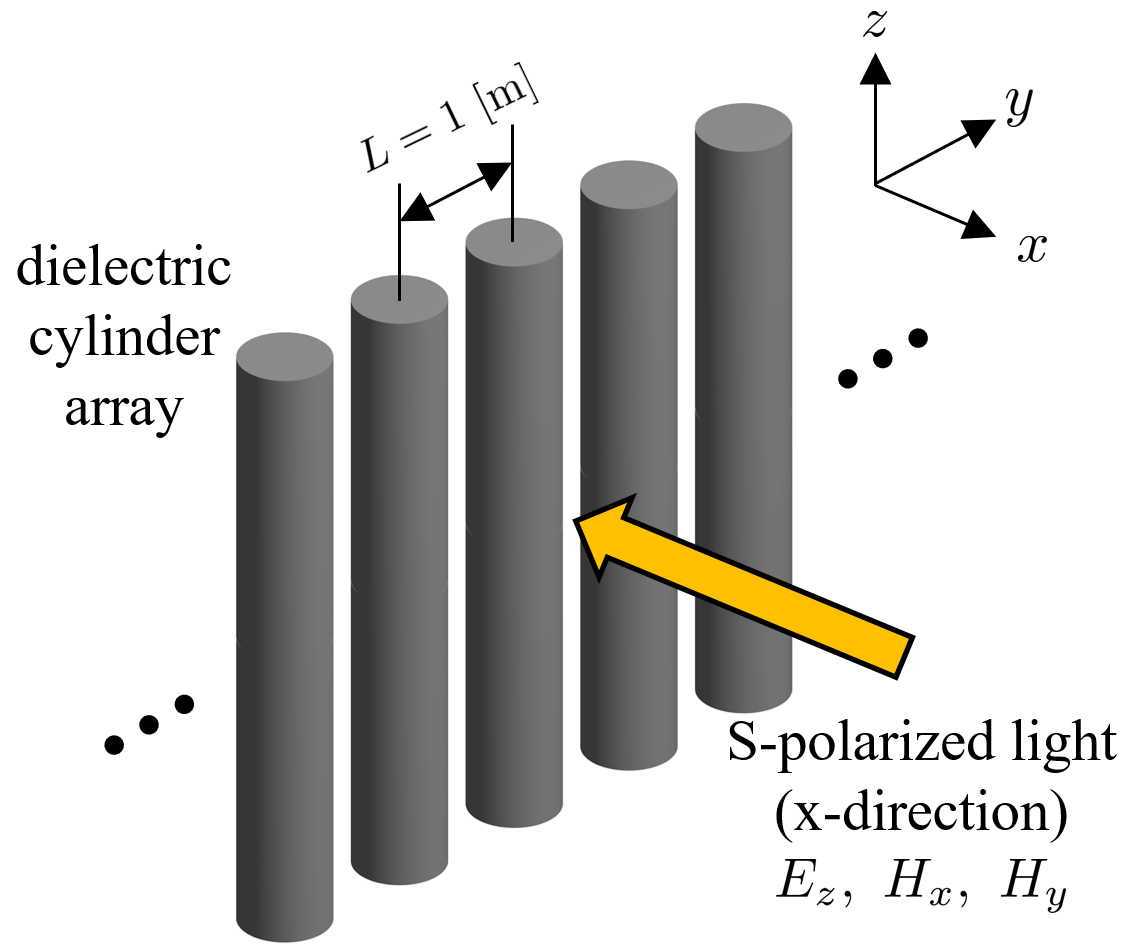}
\caption{Schematic of a periodic dielectric cylinder array.}
\label{fig:validation_geometry}
\end{figure}

To validate the developed two-dimensional FDTD algorithm, we computed the reflectance and transmittance for an S-polarized plane wave incident on a periodic dielectric cylinder array, as illustrated in Fig.~\ref{fig:validation_geometry}.  
The resulting spectra were compared against the theoretical predictions reported in~\cite{hu2018resonances}.

\begin{figure}[ht]
\centering
\includegraphics[width=0.5\linewidth]{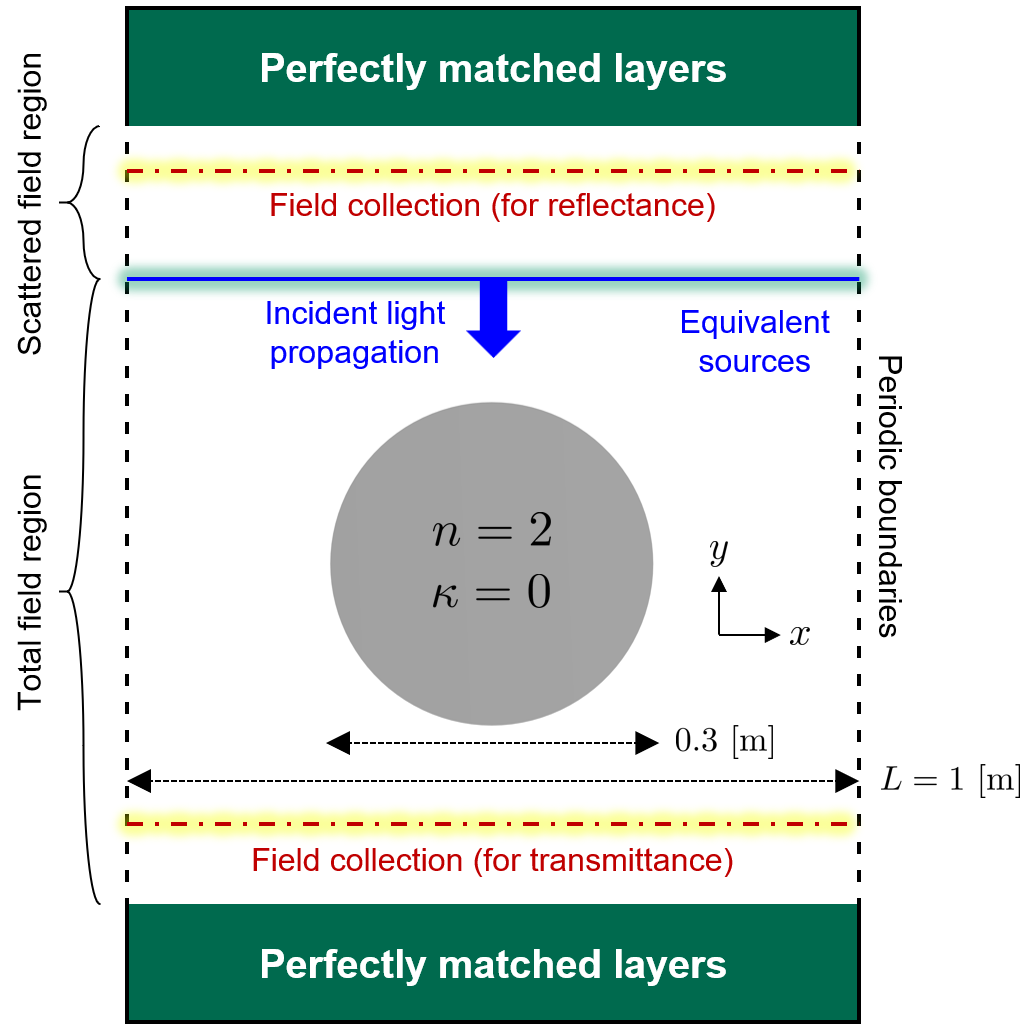}
\caption{FDTD simulation setup for the periodic dielectric array under normal incidence of an S-polarized wave.}
\label{fig:fdtd_sim_setup}
\end{figure}

Because the dielectric array extends infinitely in the transverse direction, only a single unit cell was simulated by truncating one period of the array and modeling its cross-section in the $xy$-plane.  
The computational domain correspondingly represents this unit cell, with periodic boundary conditions applied along the $y$-direction and PML boundaries applied along the $x$-direction, as shown in Fig.~\ref{fig:fdtd_sim_setup}.  

The array period was set to $L = 1~\mathrm{m}$, the cylinder diameter to $D = 0.3~\mathrm{m}$, and the refractive index of the dielectric cylinders to $n = 2$.  
An S-polarized plane wave was normally incident onto the structure, and the wavelength was swept from $1~\mathrm{m}$ to $2~\mathrm{m}$.  
The FDTD simulation parameters are summarized in Table~\ref{tab:validation_sim_param}.

\begin{table}[h]
\centering
\caption{FDTD simulation parameters for the dielectric array.}
\label{tab:validation_sim_param}
\begin{tabular}{l c}
\toprule
\textbf{Grid points in $x$ ($N_{g,x}$)} & 101 \\
\textbf{Spatial step $h$} & 0.01 m \\
\textbf{Grid points in $y$ ($N_{g,y}$)} & 897 \\
\textbf{Time step $\Delta t$} & 23.57 ps \\
\bottomrule
\end{tabular}
\end{table}

\begin{figure}[ht]
\centering
\includegraphics[width=0.5\linewidth]{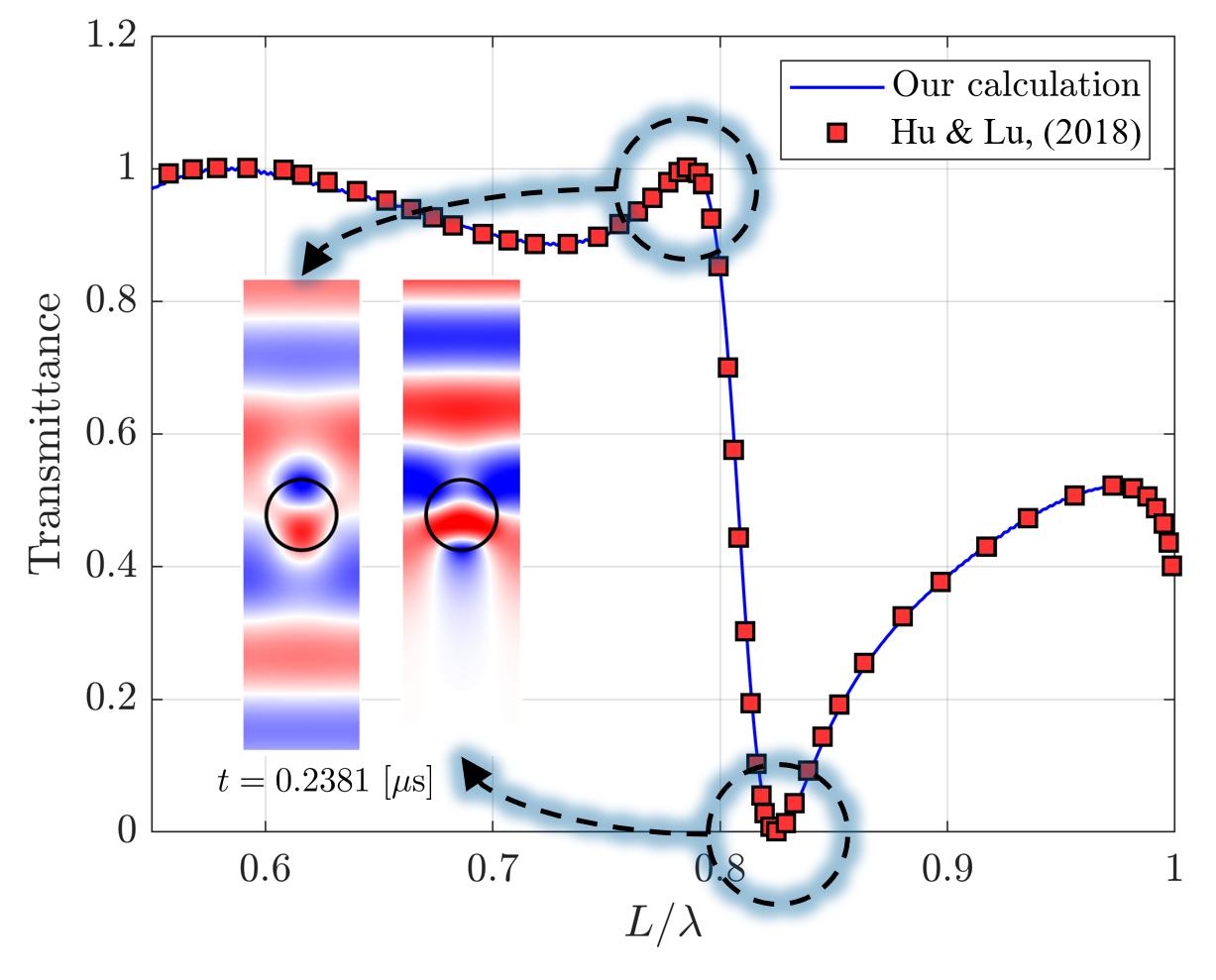}
\caption{Comparison of FDTD-computed spectral transmittance with the theoretical results of~\cite{hu2018resonances}.  
The left insets show snapshots of the electric field ($E_z$) distribution at $t = 0.2381~\mu\text{s}$ for $L/\lambda \approx 0.82$ and $L/\lambda \approx 0.79$.}
\label{fig:validation_result}
\end{figure}

Fig.~\ref{fig:validation_result} presents the transmittance as a function of the normalized electrical length $L/\lambda$.  
The FDTD results exhibit excellent agreement with the theoretical curve from~\cite{hu2018resonances}, thereby confirming the accuracy and reliability of the developed 2D FDTD formulation.

In particular, when $L/\lambda \approx 0.82$, the transmittance sharply drops to nearly zero due to the excitation of resonant modes supported by the periodic structure.  
These resonances correspond to \textit{bound states in the continuum} (BICs), in which incoming energy is trapped within the array and cannot propagate through~\cite{Friedrich1985BIC}.  
For $L/\lambda < 0.8$, these BIC modes are not excited, and most of the incoming wave transmits through the array with minimal attenuation.  
The inset field distributions in Fig.~\ref{fig:validation_result} clearly demonstrate the contrasting behaviors of strong resonance trapping and efficient transmission.

\section{Results}

In this section, we investigate two scenarios.  
First, the spectral reflectance and transmittance of a healthy plant leaf are computed using the proposed FDTD-based framework and compared with those obtained from the PROSPECT-PRO model~\cite{feret2021prospect}.  
By analyzing both dicot and monocot samples, we show that the VIS--NIR optical responses exhibit distinct characteristics depending on the internal anatomical structure of the leaf.
Second, we examine a dicot leaf exhibiting early-stage fungal infection on its surface.  
The spectral reflectance and transmittance of the diseased region are evaluated using the same FDTD simulations and subsequently compared with those of the healthy leaf.  
The observed spectral changes are further discussed in connection with previously reported biophysical interpretations of pathogen-induced tissue alterations.

\begin{figure}[htbp]
    \centering
    \subfloat[\centering Dicot]{
        \includegraphics[width=0.6\textwidth]{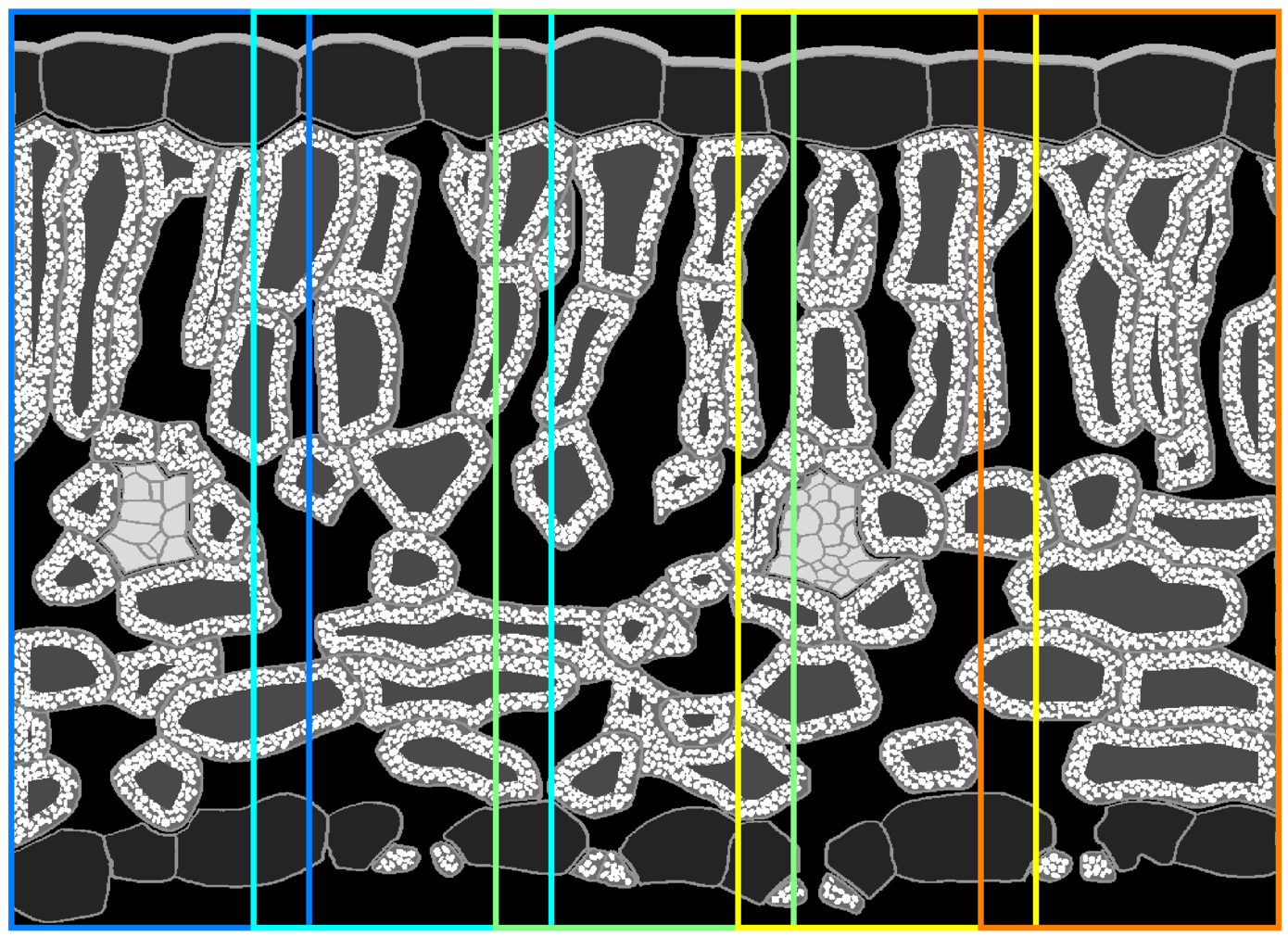}
        \label{fig:2D_FDTD_Yee_grid_global_sub1}
    }\\[2mm]
    \subfloat[\centering Monocot]{
        \includegraphics[width=0.85\textwidth]{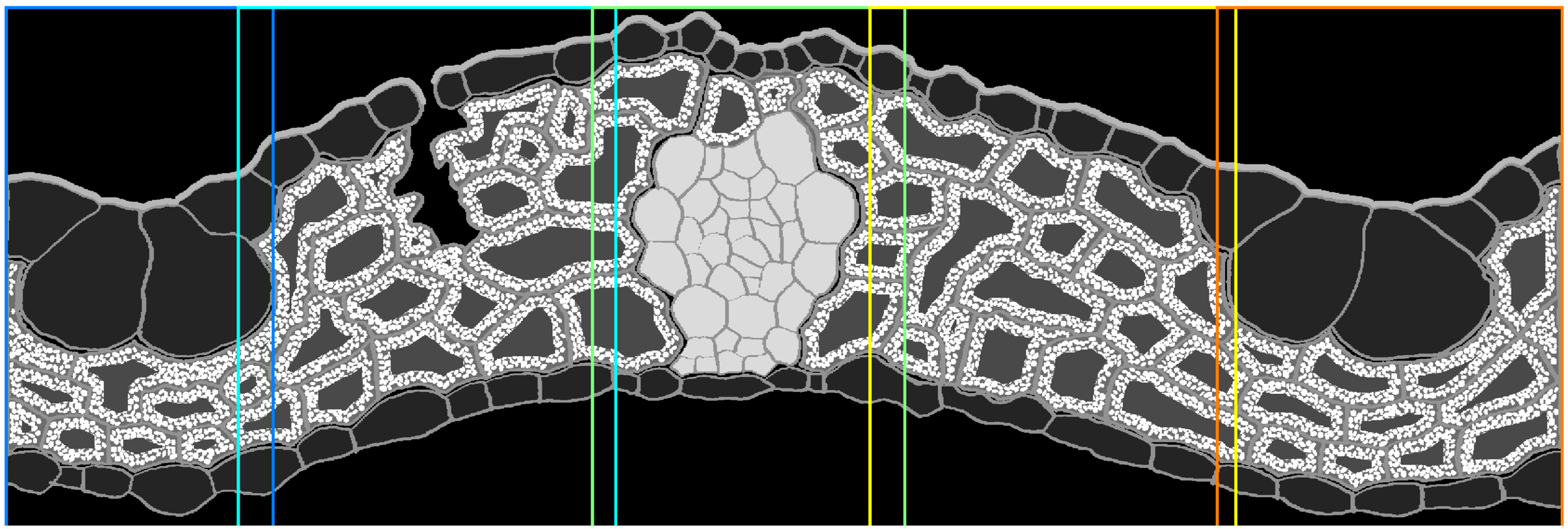}
        \label{fig:2D_FDTD_Yee_grid_global_sub2}
    }
    \caption{Two-dimensional FDTD Yee grids for the cross-sections of (a) dicot and (b) monocot plant leaves, where different colors indicate segmented tissue regions.}
    \label{fig:2D_FDTD_Yee_grid_global}
\end{figure}

The physical dimensions of the plant leaf samples and the corresponding FDTD simulation parameters are summarized as follows.
For the dicot sample, shown in Fig.~\ref{fig:2D_FDTD_Yee_grid_global_sub1}, the leaf thickness was approximately $140~\mu\mathrm{m}$ and the width was $200~\mu\mathrm{m}$.
\begin{figure}[htbp]
    \centering
    \subfloat[\centering Dicot]{
        \includegraphics[width=0.6\textwidth]{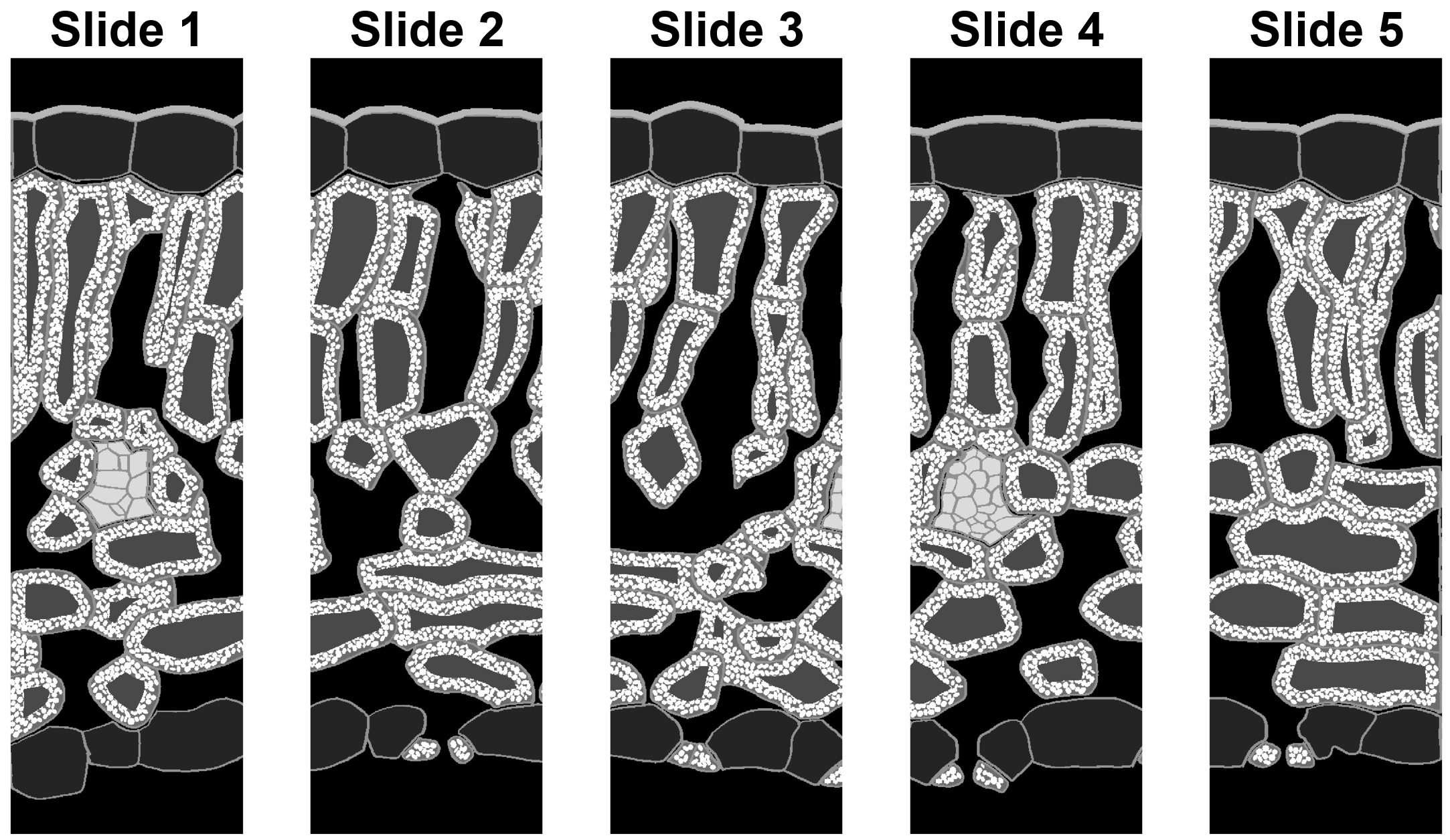}
        \label{fig:2D_FDTD_Yee_grid_slide_sub1}
    }\\[2mm]
    \subfloat[\centering Monocot]{
        \includegraphics[width=0.85\textwidth]{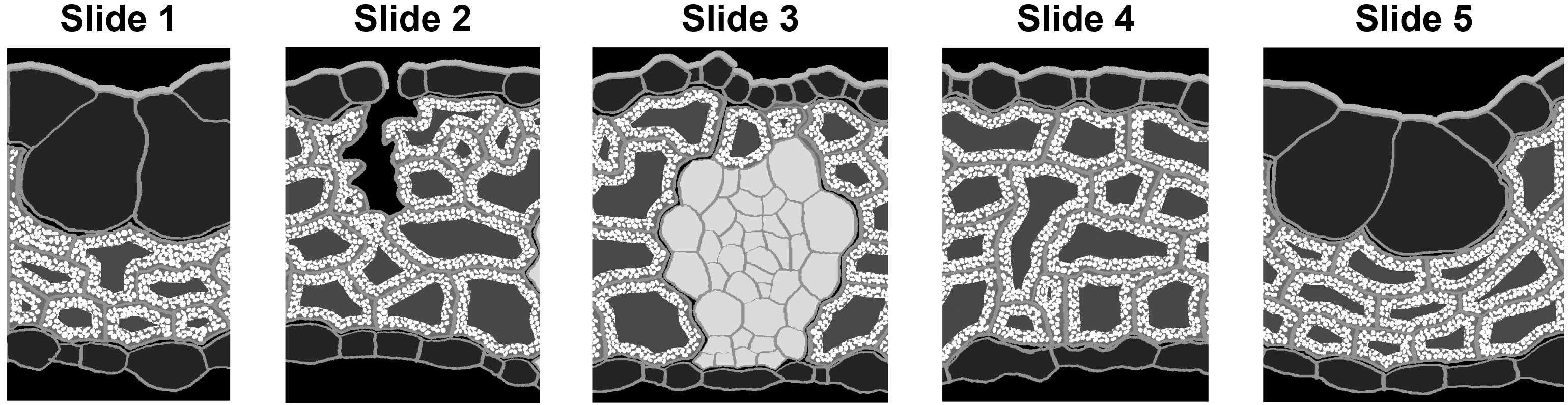}
        \label{fig:2D_FDTD_Yee_grid_slide_sub2}
    }
    \caption{Division of the full leaf cross-section into five horizontal segments for independent FDTD simulations.}
    \label{fig:2D_FDTD_Yee_grid_slide}
\end{figure}
The wavelength range of interest spans $400$--$2,500~\mathrm{nm}$.  
Based on the shortest wavelength, the computational domain corresponds to an electrical size of approximately $350 \times 500$.  
To ensure numerical stability and accuracy, the mesh size was chosen considering both the effective wavelength inside tissue,
\[
\lambda_{\mathrm{eff}} = \frac{\lambda_0}{n},
\qquad
\delta = \frac{\lambda}{2\pi\kappa},
\]
where $\delta$ is the skin depth.  
Using the minimum values over the operating band, a cell-per-wavelength (CPW) of approximately $11.2$ was maintained:
\[
\lambda_{\min} = \frac{\lambda_0}{n_{\mathrm{cell~wall}}} \approx 263.16~\mathrm{nm},
\qquad
h = \frac{\lambda_{\min}}{\mathrm{CPW}} \approx 23.41~\mathrm{nm}.
\]
The time-step interval was then set by the CFL condition:
\[
\Delta t \approx \frac{h}{c\sqrt{2}} = 0.0552~\mathrm{fs}.
\]
A total of $N_t = 55{,}000$ time steps were executed, corresponding to a physical duration of approximately $3~\mathrm{ps}$.  
This is equivalent to about 2,277 optical periods at $\lambda=400~\mathrm{nm}$ and 364 periods at $\lambda=2,500~\mathrm{nm}$, which is sufficient for the fields to reach steady state after all scattering processes.
For the dicot sample, the resulting discretized grid contains approximately $8,511 \times 6,689$ cells (including PMLs), corresponding to about $5.7\times10^7$ unknowns.  
To mitigate this computational burden, the dicot leaf image was divided into five horizontal slices (Fig.~\ref{fig:2D_FDTD_Yee_grid_slide}).  
Independent FDTD simulations were carried out for each slice, and the overall reflectance and transmittance were obtained by averaging the slice-wise results.
A similar strategy was applied to the monocot leaf (Fig.~\ref{fig:2D_FDTD_Yee_grid_global_sub2}).  
For this case, the simulation parameters were $h = 23.15~\mathrm{nm}$ and $\Delta t = h/(c\sqrt{2}) \approx 0.0546~\mathrm{fs}$, and the reconstructed grid comprised $13{,}538 \times 4{,}461$ cells.  
Since the monocot leaf exhibits a slight global curvature, each slice was appropriately rotated to model normal incidence accurately.  
Independent simulations were again performed for the five slices, and the final spectra were obtained by averaging the reflectance and transmittance over all sections.

\subsection{Simulation Results for Healthy Dicot and Monocot Plant Leaves}
\begin{figure}[htbp]
    \centering
    \subfloat[\centering Dicot]{%
        \includegraphics[width=0.75\textwidth]{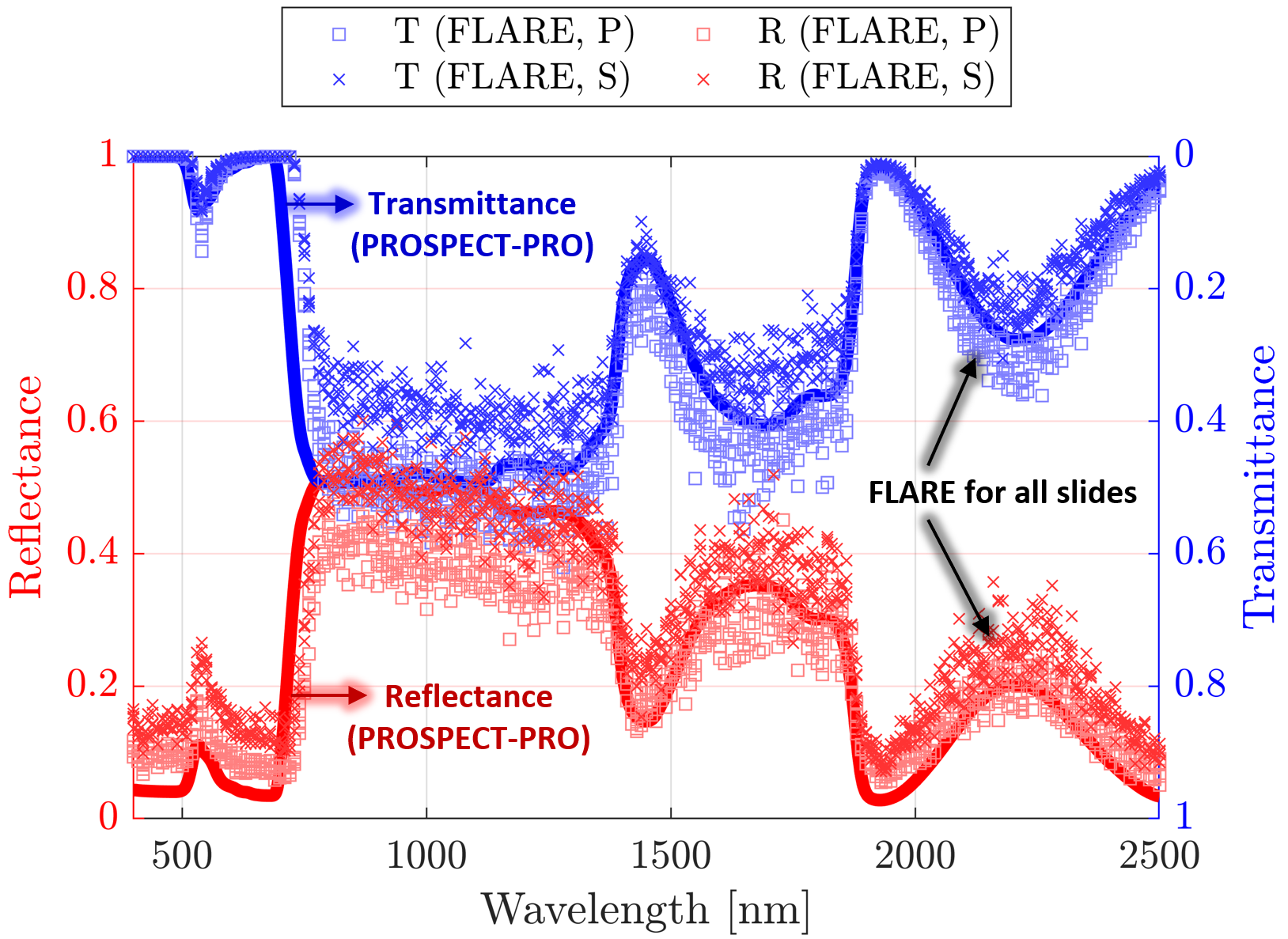}
        \label{fig:spectral_R_T_healthy_sub1}
    }\\[4pt]
    \subfloat[\centering Monocot]{%
        \includegraphics[width=0.75\textwidth]{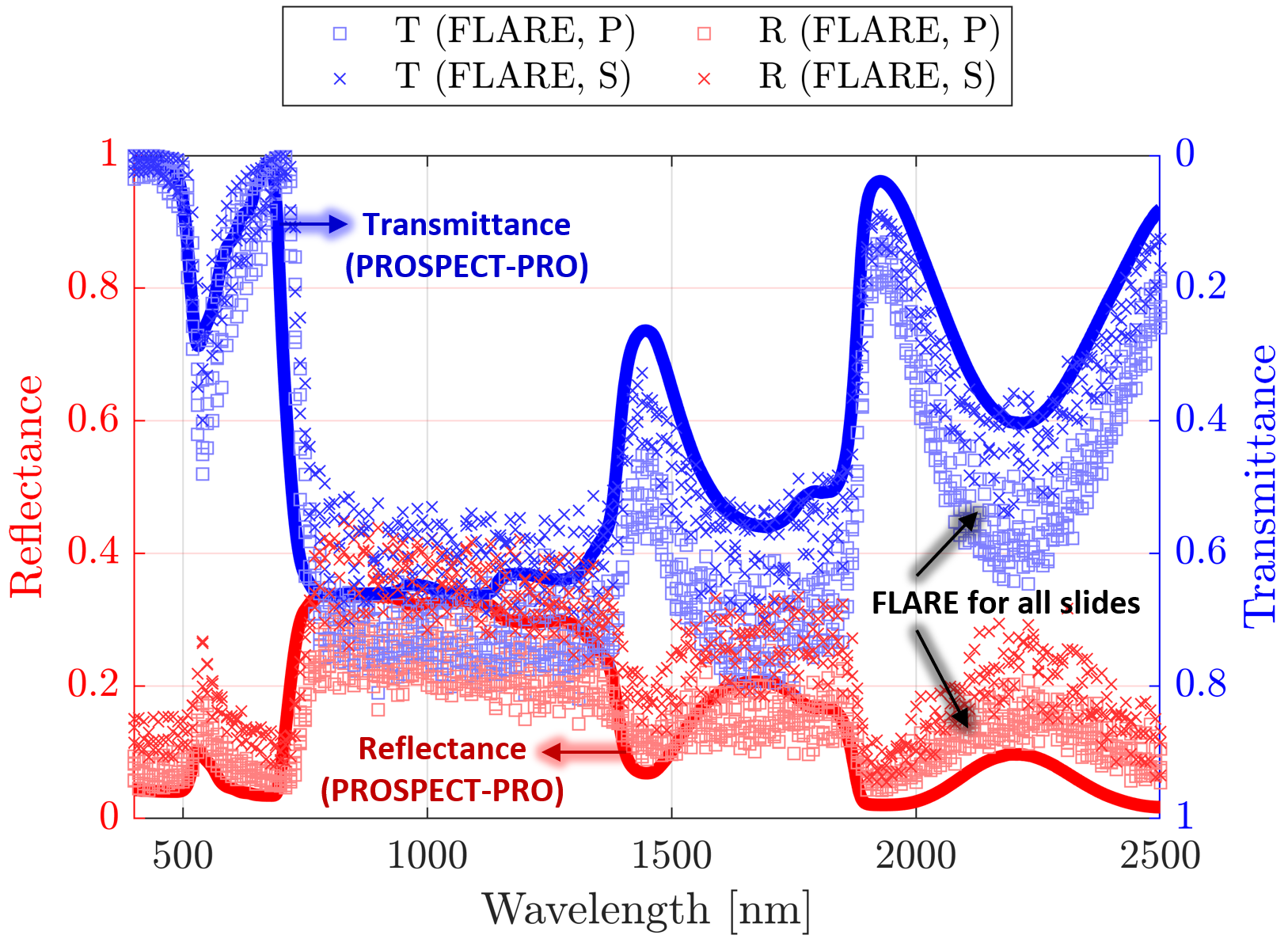}
        \label{fig:spectral_R_T_healthy_sub2}
    }
    \caption{Spectral reflectance and transmittance of (a) dicot and (b) monocot plant leaves obtained from FDTD simulations for each slide and polarization, compared with the PROSPECT-PRO model.}
    \label{fig:spectral_R_T_healthy}
\end{figure}

The FDTD-simulated spectral reflectance and transmittance of healthy dicot and monocot leaves are shown in Fig.~\ref{fig:spectral_R_T_healthy}.  
The results for each slide and both polarizations are plotted together with the PROSPECT-PRO model predictions for comparison.
Overall, the FDTD simulations exhibit excellent agreement with the PROSPECT-PRO predictions, successfully capturing the key optical responses arising from tissue microstructure, pigment content, and internal scattering.  
This demonstrates that the proposed microstructure-resolved FDTD framework accurately reproduces the wavelength-dependent optical behavior of healthy leaves.
Table~\ref{tab:dicot_mono_PROSPECT_PRO} summarizes the input parameters used for running the PROSPECT-PRO model.  
Here, $C_{ab}$, $C_{car}$, $C_{wat}$, and $N_{sca}$ denote the concentrations of chlorophyll~a+b, carotenoids, water content, and the scattering extent, respectively~\cite{feret2021prospect}.

\begin{table}[h!]
\centering
\caption{PROSPECT-PRO input parameters for dicot and monocot samples.}
\label{tab:dicot_mono_PROSPECT_PRO}
\begin{tabular}{lcc}
\hline
\textbf{Parameter} & \textbf{Dicot} & \textbf{Monocot} \\
\hline
$C_{ab}$   & 79.54  & 42.22 \\
$C_{car}$  & 11.23  & 5.04 \\
$C_{wat}$  & 0.0015 & 0.0015 \\
$N_{sca}$  & 1.65   & 0.85 \\
\hline
\end{tabular}
\end{table}

As shown in Fig.~\ref{fig:spectral_R_T_healthy}, the spectral behavior of both dicot and monocot leaves is governed by the interplay between pigment absorption and internal tissue morphology.  
In the VIS (400--750\,nm), red and blue wavelengths are strongly attenuated because chlorophylls absorb efficiently in these bands, whereas green light exhibits comparatively higher reflectance and transmittance due to weak pigment absorption.  
Monocot leaves transmit more green light than dicot leaves, primarily because they are thinner and contain a smaller total chloroplast volume.
Internal structural morphology also plays a key role in shaping the scattering characteristics.  
Dicot leaves possess irregular, highly heterogeneous cellular arrangements, including a spongy mesophyll layer filled with numerous irregularly shaped air gaps.  
These structural features induce multiple random scattering events and frequent internal reflections---often near the critical angle---which increase the effective optical path length and enhance absorption.  
In contrast, monocot leaves exhibit more regular and vertically aligned mesophyll structures with few air gaps, causing light to propagate predominantly along near-normal directions with significantly fewer internal reflections.
As a result, monocots generally show higher transmittance and lower reflectance than dicots.
In the NIR region (750--1,400\,nm), the sum of reflectance and transmittance approaches unity for both leaf types, indicating minimal absorption.  
This aligns with the fact that pigment absorption becomes negligible beyond 750\,nm; hence, spectral variations in this region are mainly governed by tissue geometry rather than biochemical composition.
A comparison of the two leaf types reveals that dicots exhibit nearly balanced reflectance and transmittance (approximately 50\% each), whereas monocots show transmittance around 65\% and reflectance around 35\%.  
Again, these differences arise from the more ordered internal architecture and smaller air cavities in monocots, which reduce random scattering and internal trapping of light.  
The PROSPECT-PRO results consistently reproduce these trends, further validating the FDTD predictions.
Beyond 1,400\,nm, three prominent absorption dips appear near 1,450\,nm, 1,900\,nm, and 2,500\,nm, corresponding to well-known water absorption bands.  
Despite the presence of strong water absorption, monocot leaves still exhibit comparatively high overall transmittance owing to their reduced thickness and orderly tissue structure.
Because plant cells---particularly the cytoplasm and vacuoles---contain substantial water, these spectral features are accurately reproduced in both the FDTD and PROSPECT-PRO simulations.  
However, the FDTD approach captures the scattering and absorption behavior directly from first principles, whereas PROSPECT-PRO cannot resolve these mechanisms explicitly and instead controls their net effect through the phenomenological scattering parameter $N_{\mathrm{sca}}$.

\begin{figure}[htbp]
    \centering
    \subfloat[\centering Dicot (TE\textsubscript{z} polarization)]{%
        \includegraphics[width=0.45\textwidth]{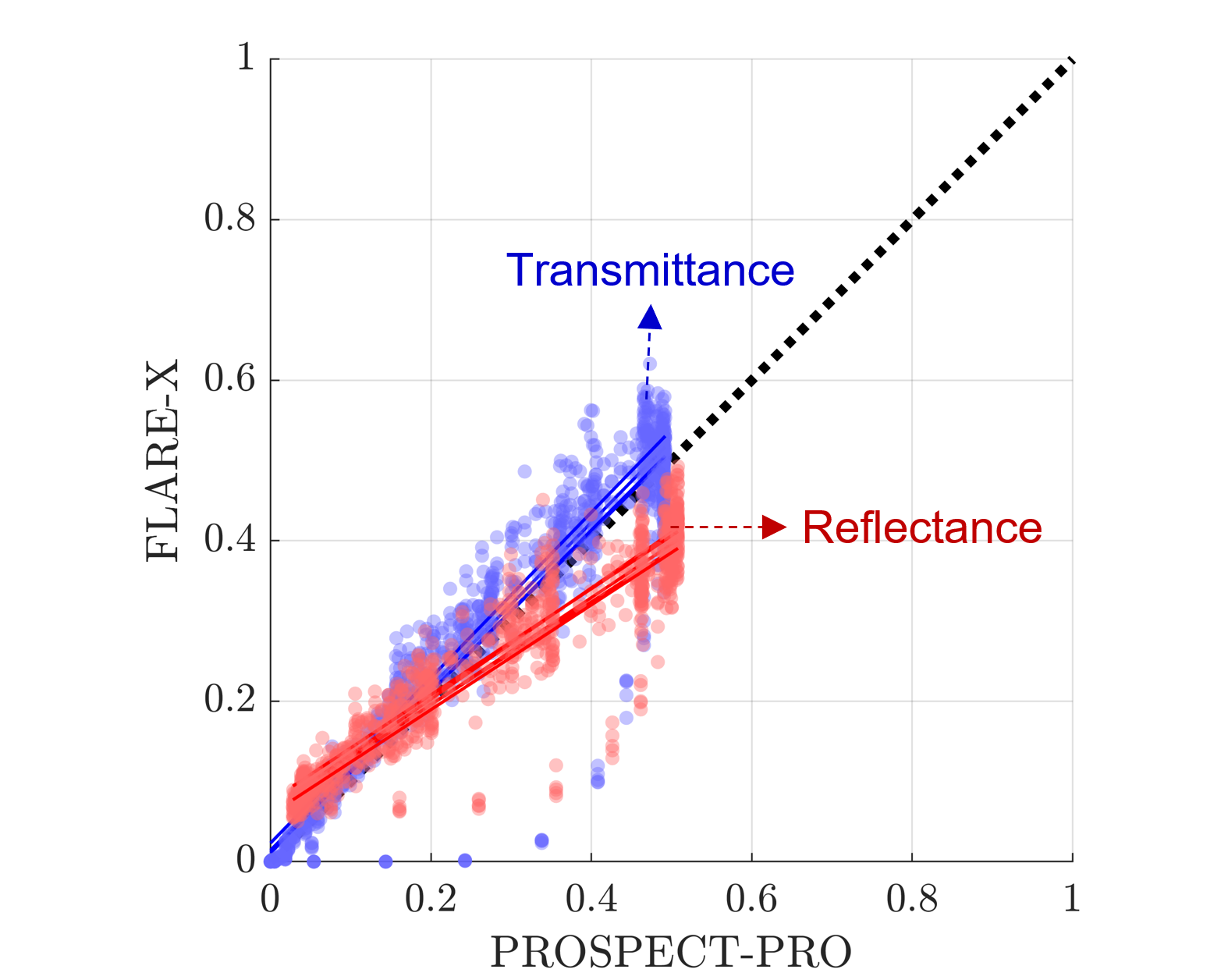}
        \label{fig:sta_healthy1_sub1}
    }
    \subfloat[\centering Dicot (TM\textsubscript{z} polarization)]{%
        \includegraphics[width=0.45\textwidth]{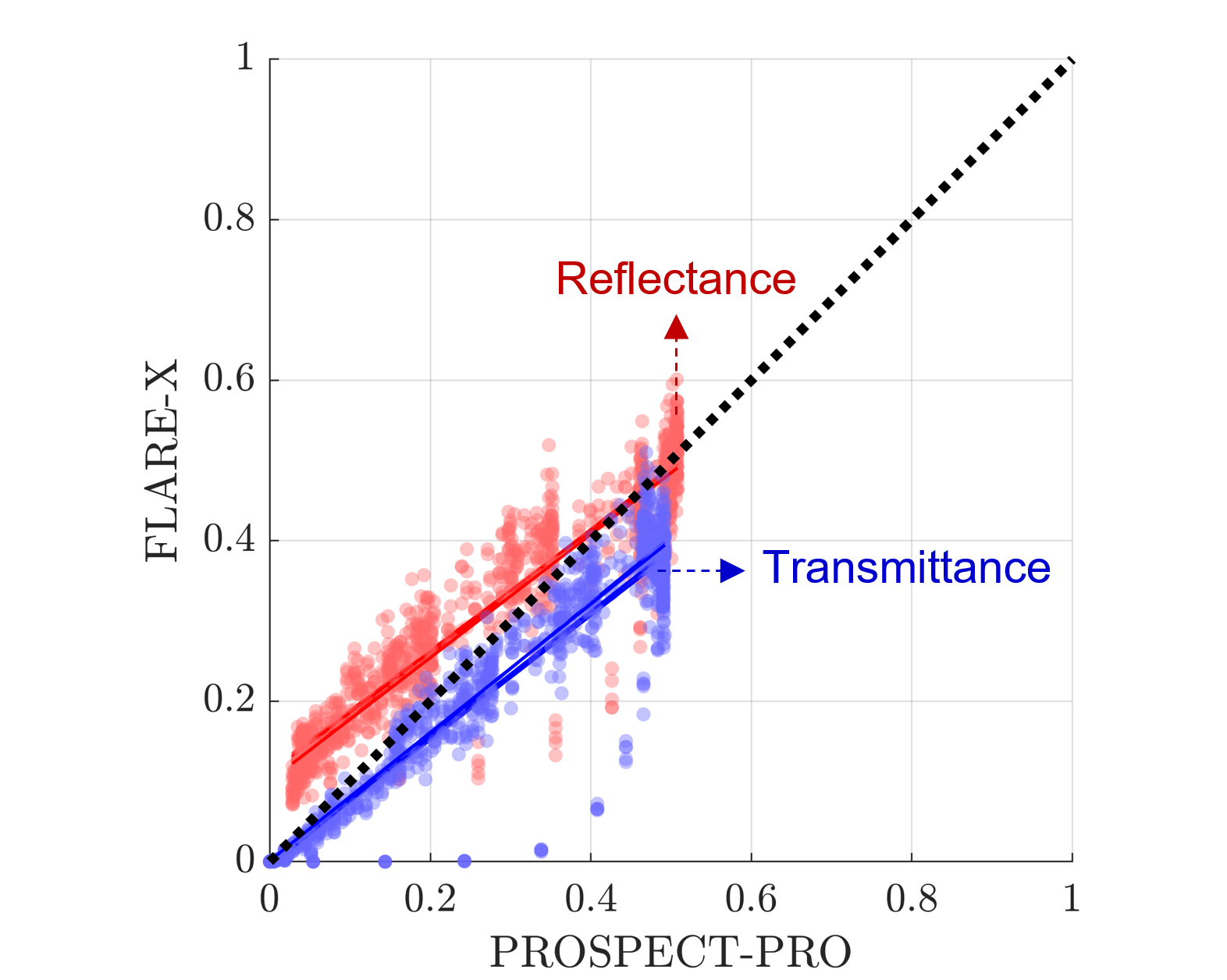}
        \label{fig:sta_healthy1_sub2}
    }
    \caption{Regression analysis between the FDTD and PROSPECT-PRO results for the dicot leaf.}
    \label{fig:sta_healthy1}
\end{figure}

\begin{figure}[htbp]
    \centering
    \subfloat[\centering Monocot (TE\textsubscript{z} polarization)]{%
        \includegraphics[width=0.45\textwidth]{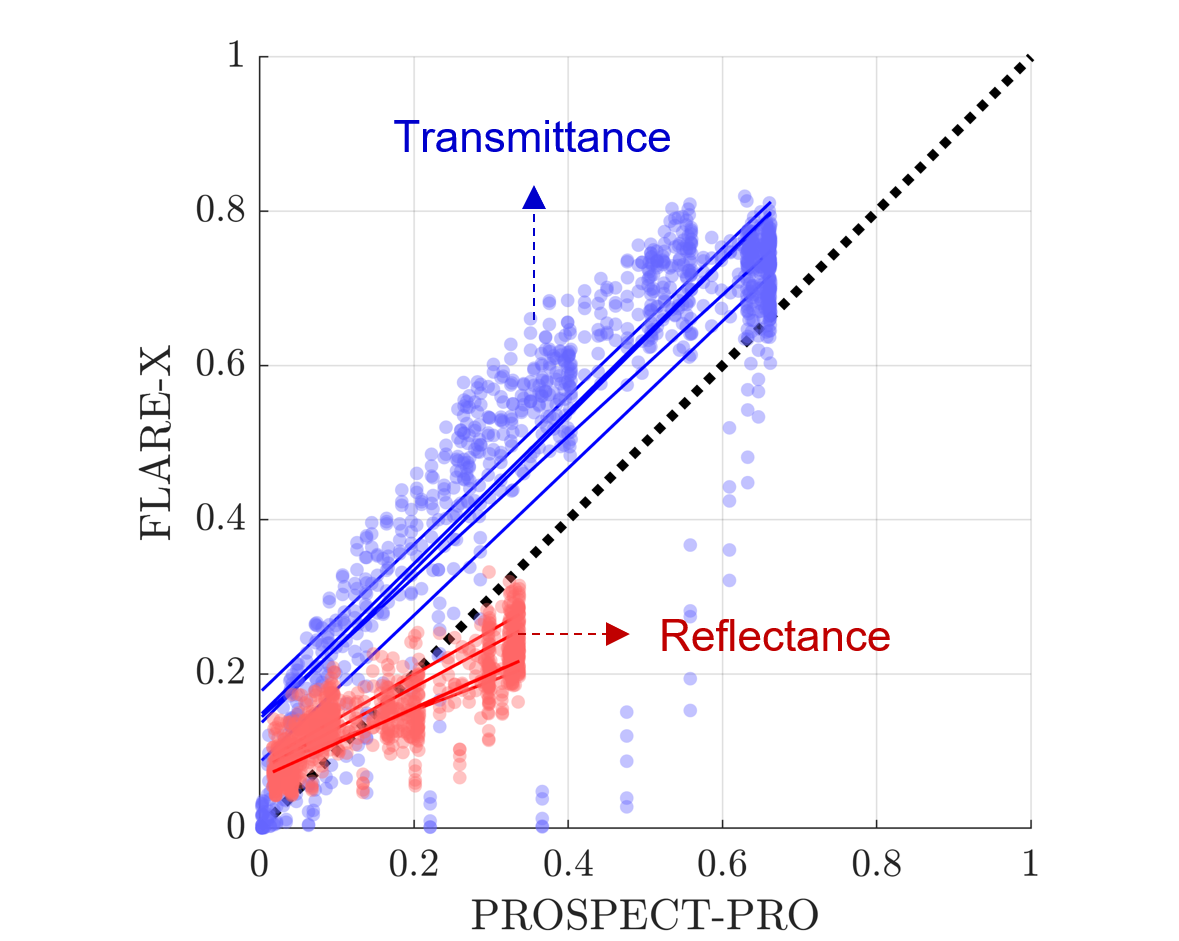}
        \label{fig:sta_healthy2_sub1}
    }
    \subfloat[\centering Monocot (TM\textsubscript{z} polarization)]{%
        \includegraphics[width=0.45\textwidth]{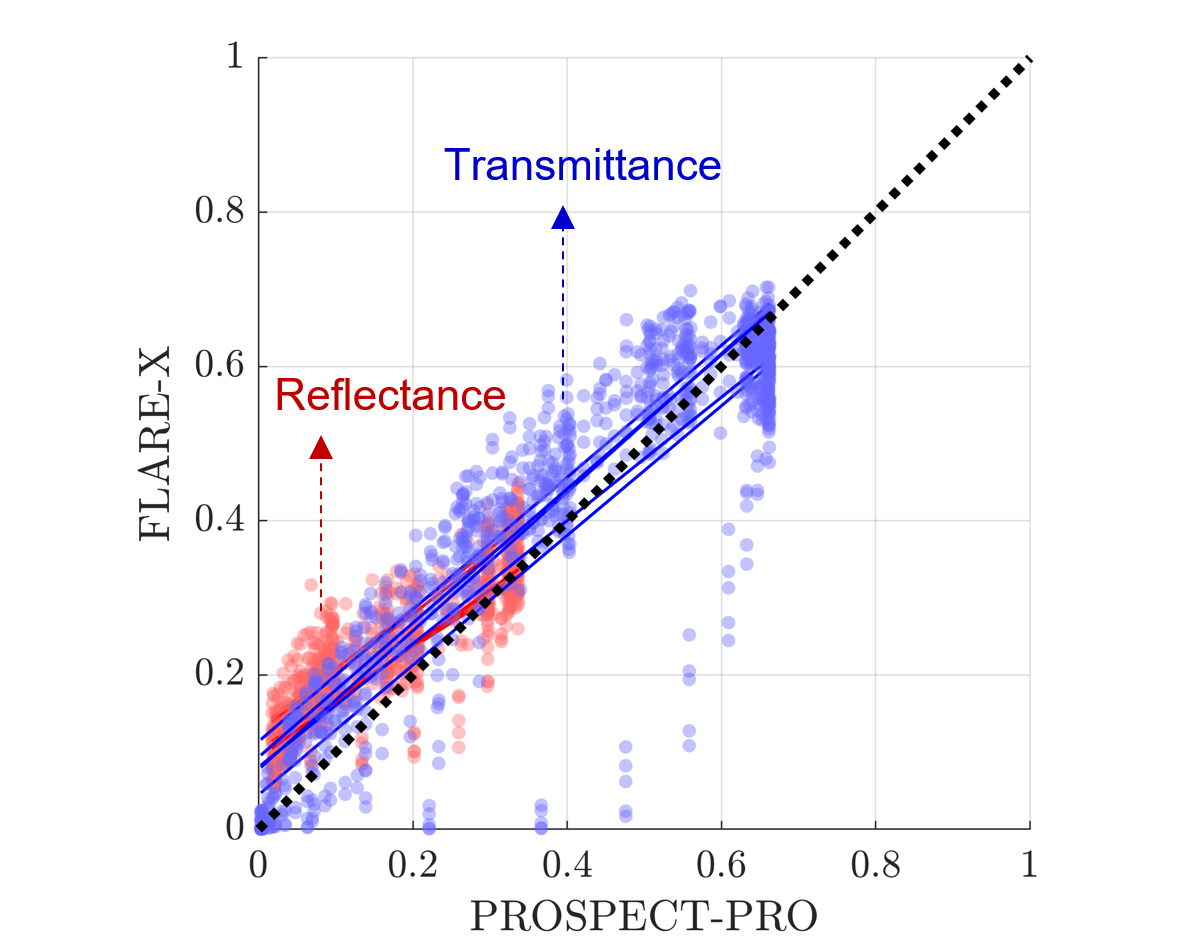}
        \label{fig:sta_healthy2_sub2}
    }
    \caption{Regression analysis between the FDTD and PROSPECT-PRO results for the monocot leaf.}
    \label{fig:sta_healthy2}
\end{figure}

\begin{figure}[htbp]
    \centering
    \subfloat[\centering Dicot]{%
        \includegraphics[width=0.45\textwidth]{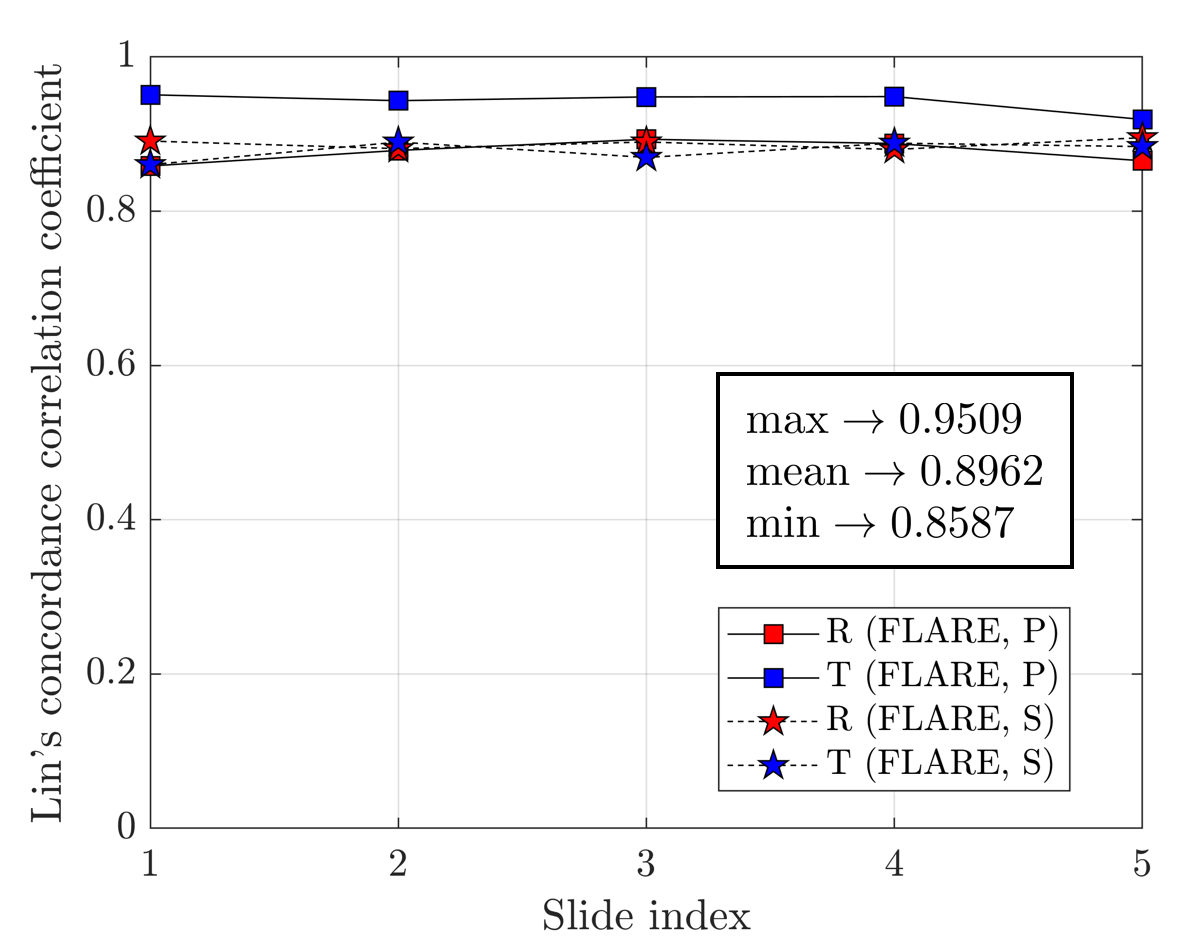}
        \label{fig:sta_healthy3_sub1}
    }
    \subfloat[\centering Monocot]{%
        \includegraphics[width=0.45\textwidth]{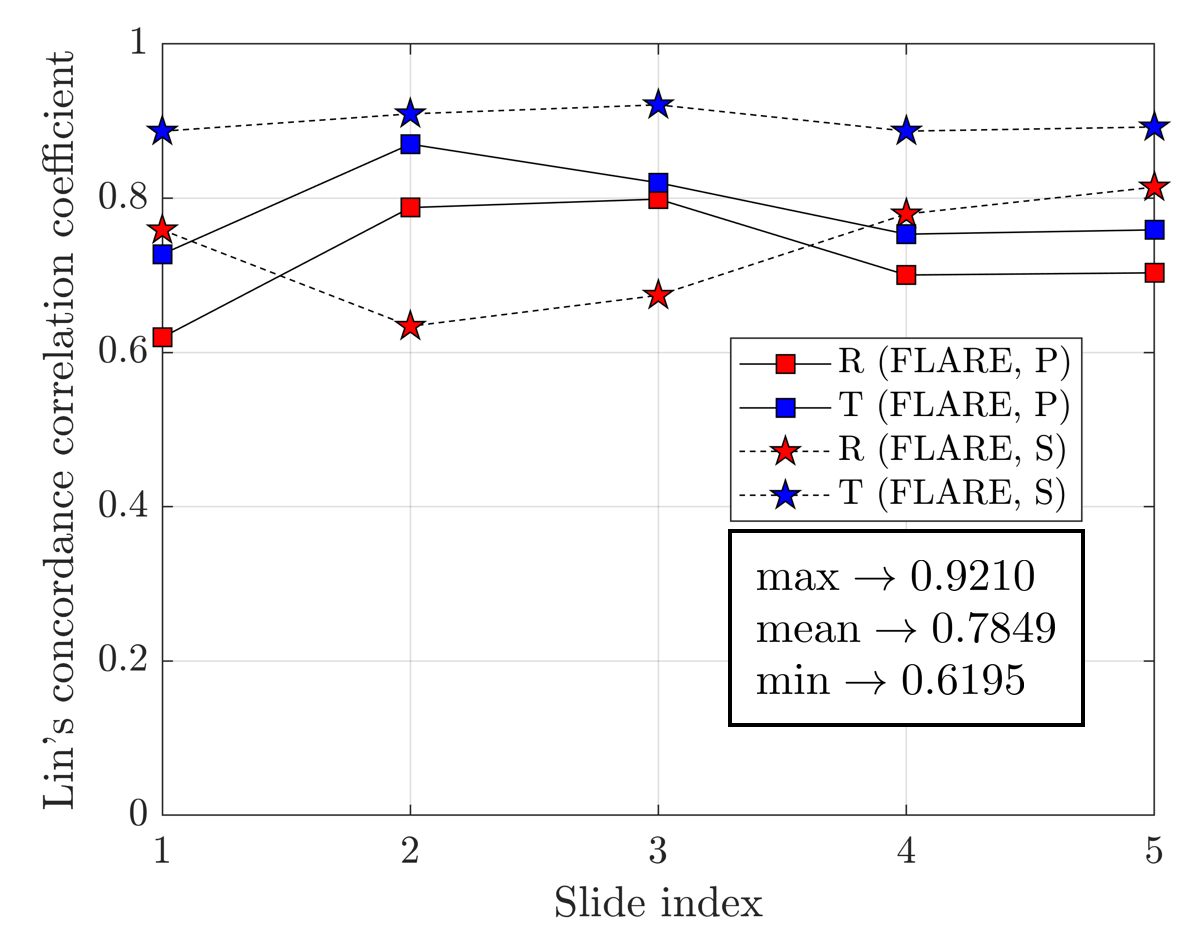}
        \label{fig:sta_healthy3_sub2}
    }
    \caption{Lin’s concordance correlation coefficient (CCC) analysis between the FDTD and PROSPECT-PRO results for dicot and monocot leaves, showing the CCC values for each slide and polarization.}
    \label{fig:sta_healthy3}
\end{figure}

To quantitatively assess the agreement between the FDTD simulations and the PROSPECT-PRO model, regression analyses were performed for each slide and polarization, as shown in Figs.~\ref{fig:sta_healthy1}--\ref{fig:sta_healthy3}.  
For the dicot sample, all datasets exhibit strong linear correlations, confirming the validity and robustness of the proposed full-wave optical modeling framework.  
In the monocot case, slight deviations appear in the P-polarized reflectance, which may stem from the more anisotropic and directionally aligned internal structure of monocot tissues, resulting in polarization-dependent scattering behavior.

The Lin’s concordance correlation coefficient (CCC) was computed for all slides and both polarizations.  
For the dicot samples, all CCC values exceeded 0.8587, with a maximum of 0.9509 and an average of 0.8962, indicating consistently strong agreement between the FDTD simulations and the PROSPECT-PRO predictions.  
For the monocot samples, the mean CCC was 0.7849, with maximum and minimum values of 0.9210 and 0.6195, respectively, reflecting slightly larger variability---likely stemming from the stronger structural anisotropy inherent in monocot tissues.  
The detailed Pearson correlation coefficients ($r$-values) and corresponding significance levels ($p$-values) are provided in Tables~\ref{tab:correlation_rotated1} and~\ref{tab:correlation_rotated2}.
\begin{table}[htbp]
\centering
\renewcommand{\arraystretch}{1.2}
\setlength{\tabcolsep}{6pt}
\begin{tabular}{c|cc|cc}
\hline
\multirow{2}{*}{\textbf{Slide}} & 
\multicolumn{2}{c|}{\textbf{Reflectance}} & 
\multicolumn{2}{c}{\textbf{Transmittance}} \\
\cline{2-5}
 & r-val. & p-val. & r-val. & p-val. \\
\hline
S1 (P) & 0.9326 & 1.7e-94 & 0.9540 & 2.1e-111 \\
S1 (S) & 0.9434 & 3.7e-102 & 0.9476 & 1.2e-105 \\
S2 (P) & 0.9434 & 3.4e-102 & 0.9515 & 5.8e-109 \\
S2 (S) & 0.9502 & 8.5e-108 & 0.9498 & 1.8e-107 \\
S3 (P) & 0.9396 & 2.6e-99 & 0.9511 & 1.2e-108 \\
S3 (S) & 0.9423 & 2.6e-101 & 0.9474 & 1.9e-105 \\
S4 (P) & 0.9493 & 5.0e-107 & 0.9522 & 1.1e-109 \\
S4 (S) & 0.9420 & 4.3e-101 & 0.9457 & 4.9e-104 \\
S5 (P) & 0.9438 & 1.6e-102 & 0.9365 & 3.9e-97 \\
S5 (S) & 0.9454 & 9.5e-104 & 0.9439 & 1.4e-102 \\
\hline
\end{tabular}
\caption{Pearson correlation coefficients ($r$-values) and significance ($p$-values) for reflectance and transmittance in dicot leaves (slides S1-S5).}
\label{tab:correlation_rotated1}
\end{table}
\begin{table}[htbp]
\centering
\renewcommand{\arraystretch}{1.2}
\setlength{\tabcolsep}{6pt}
\begin{tabular}{c|cc|cc}
\hline
\multirow{2}{*}{\textbf{Slide}} & 
\multicolumn{2}{c|}{\textbf{Reflectance}} & 
\multicolumn{2}{c}{\textbf{Transmittance}} \\
\cline{2-5}
 & r-val. & p-val. & r-val. & p-val. \\
\hline
S1 (P) & 0.8632 & 5.5e-64 & 0.9103 & 5.1e-82 \\
S1 (S) & 0.9126 & 3.5e-83 & 0.9208 & 1.8e-87 \\
S2 (P) & 0.8706 & 2.4e-66 & 0.9092 & 1.6e-81 \\
S2 (S) & 0.8798 & 1.8e-69 & 0.9162 & 5.4e-85 \\
S3 (P) & 0.9199 & 6.1e-87 & 0.9199 & 6.3e-87 \\
S3 (S) & 0.9222 & 3.3e-88 & 0.9319 & 4.6e-94 \\
S4 (P) & 0.8894 & 5.3e-73 & 0.8827 & 1.6e-70 \\
S4 (S) & 0.9184 & 3.9e-86 & 0.8997 & 3.4e-77 \\
S5 (P) & 0.9064 & 3.5e-80 & 0.9024 & 2.1e-78 \\
S5 (S) & 0.9346 & 7.7e-96 & 0.9101 & 6.5e-82 \\
\hline
\end{tabular}
\caption{Pearson correlation coefficients ($r$-values) and significance ($p$-values) for reflectance and transmittance in monocot leaves (slides S1-S5).}
\label{tab:correlation_rotated2}
\end{table}

\begin{figure}[htbp]
\centering
\includegraphics[width=1\linewidth]{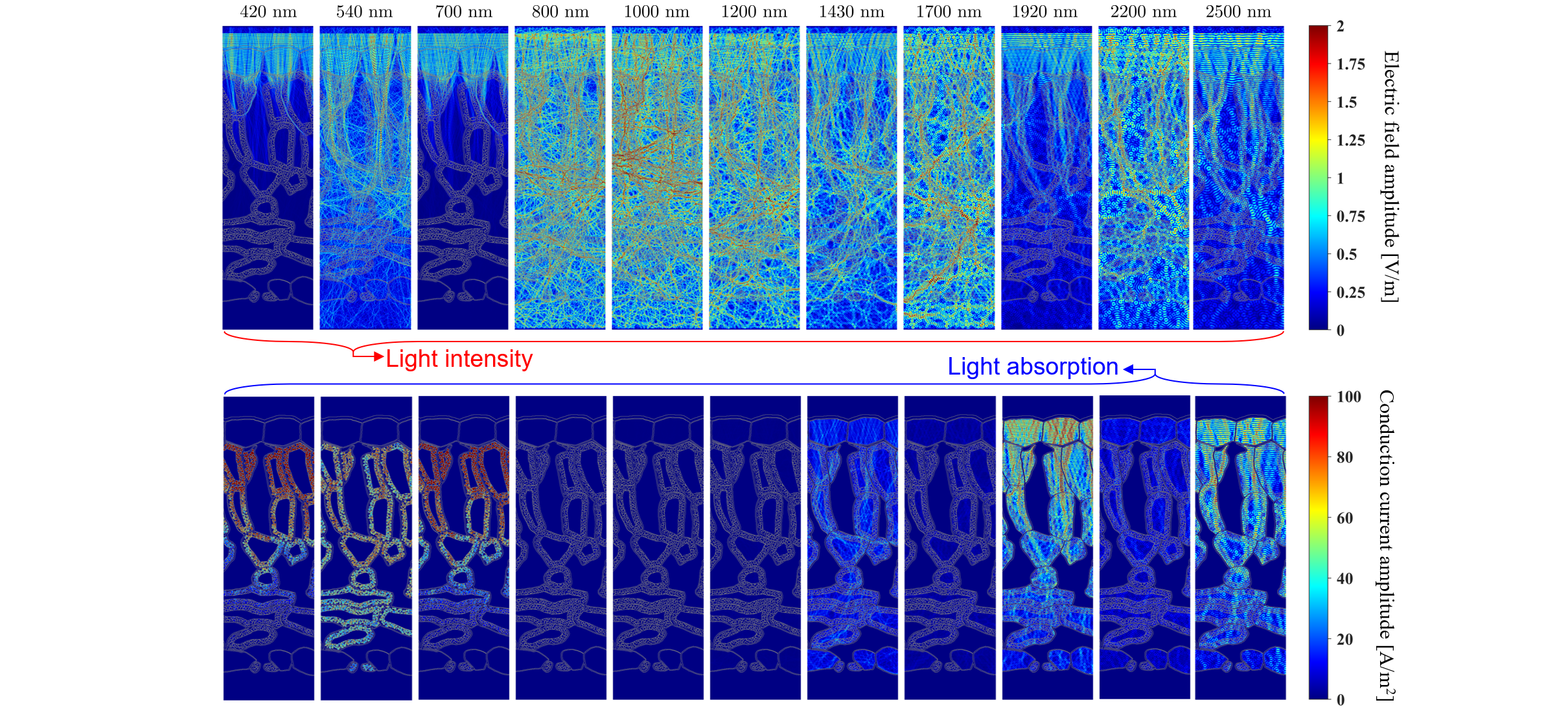}
\caption{Electric field amplitude and conduction current density distributions in the dicot leaf (slide 2) at different wavelengths.}
\label{fig:Dicot_field_healthy}
\end{figure}

\begin{figure}[htbp]
\centering
\includegraphics[width=1\linewidth]{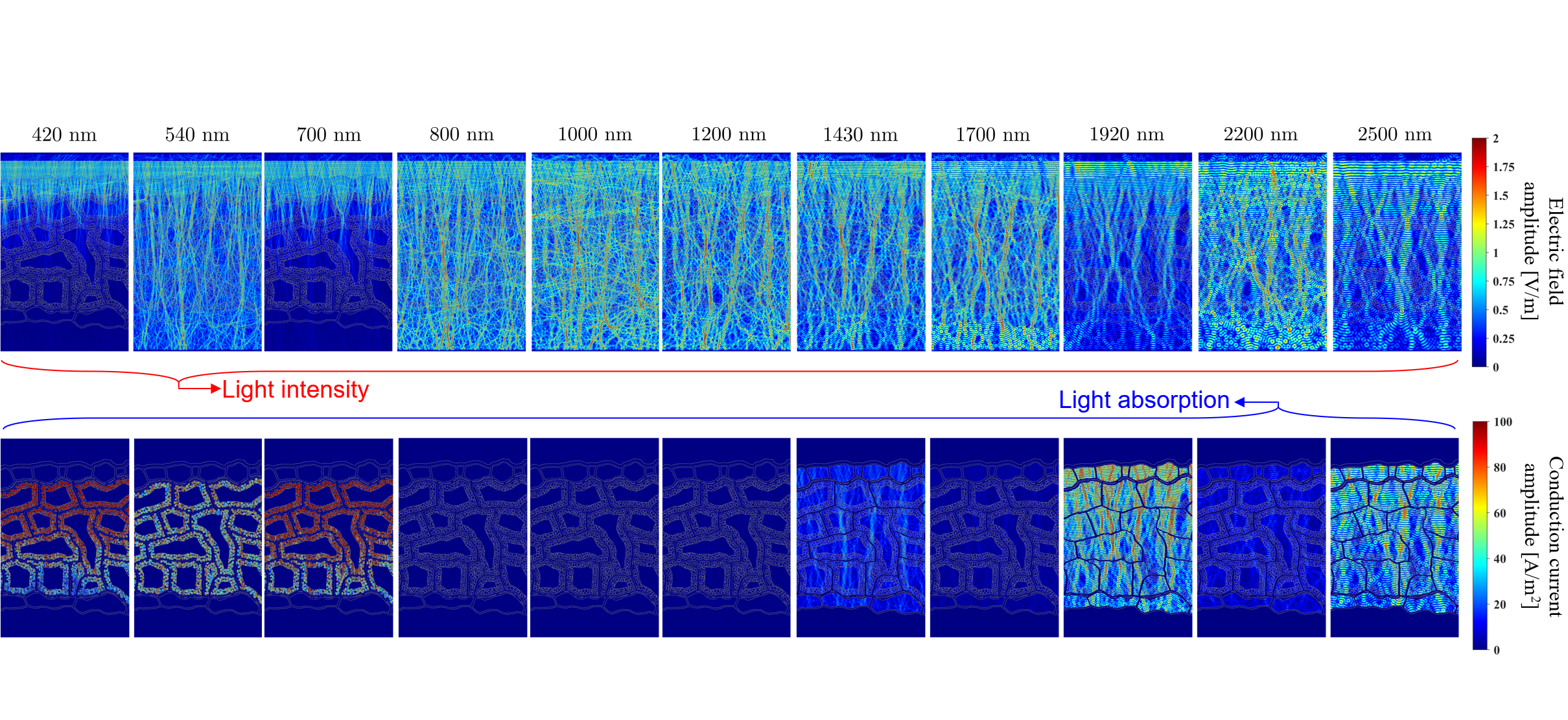}
\caption{Electric field amplitude and conduction current density distributions in the monocot leaf (slide 4) at different wavelengths.}
\label{fig:Monocot_field_healthy}
\end{figure}

Finally, Fig.~\ref{fig:Dicot_field_healthy} presents the electric field amplitude  
$|\mathbf{E}(\mathbf{r},t)|$ and the conduction current density  
$\mathbf{J}_c(\mathbf{r},t)=\sigma \mathbf{E}(\mathbf{r},t)$ for the dicot leaf under  
P-polarized illumination at steady state ($t = 55001\Delta t$).  
In the visible range, red and blue wavelengths are strongly absorbed by chloroplasts in the palisade mesophyll, as indicated by pronounced conduction-current intensities.  
Green light, by contrast, exhibits much weaker $\mathbf{J}_c$, reflecting its relatively low pigment absorption and higher transmission through the tissue.
Beyond approximately 800\,nm, the field patterns become highly irregular while  
$\mathbf{J}_c$ nearly vanishes, indicating negligible absorption in this region.  
Here, light--tissue interactions are dominated by multiple reflections and refractions at cellular boundaries.  
Due to the irregular geometry of dicot tissues---especially the large, uneven air gaps in the spongy mesophyll---scattering occurs in random directions and produces complex interference structures.  
Accordingly, the NIR reflectance and transmittance encode structural signatures of the internal cellular arrangement.
At 1,400\,nm, 1,920\,nm, and 2,500\,nm, strong conduction-current responses reappear within cytoplasmic regions, corresponding to the known water-absorption bands.
Fig.~\ref{fig:Monocot_field_healthy} shows analogous field distributions for the monocot sample. 
While the wavelength-dependent trends are qualitatively consistent with those of the dicot leaf, the field patterns in monocot tissues are notably more directional and less diffuse.  
This arises from the highly ordered, vertically aligned mesophyll structure and significantly smaller air gaps, which reduce random scattering and minimize internally reflected paths.  
As a result, a larger fraction of light exits through the lower surface, yielding higher transmittance compared to dicot leaves.

\subsection{Fungi-infected Plant Leaf}
Next, we investigate a scenario in which a necrotrophic fungal infection develops on the surface of a dicot leaf.  
Starting from the same dicot geometry used in the healthy-leaf analysis, fungal hyphae were manually drawn to emulate early-stage infection.  
Although the exact species is not specified, the modeled morphology reflects common phytopathogens such as \textit{Magnaporthe oryzae}, \textit{Colletotrichum orbiculare}, and \textit{Alternaria alternata}.  
These fungi typically form melanized appressoria capable of generating high turgor pressure and secreting cutinase and cellulase enzymes to breach the cuticle and upper epidermis.
In particular, \textit{M.~oryzae} develops a heavily melanized wall layer (30--100~nm thick), with melanin having a complex refractive index of  
$n = 1.7\text{--}1.9$ and $\kappa = 0.005\text{--}0.02$ in the VIS band.  
This high absorption reduces surface reflectance and alters local irradiance distributions.  
Accordingly, in the present simulation the hyphae were modeled with a melanin-rich outer shell and a chitin--glucan inner wall, consistent with known melanized-fungal microstructure.

Plant-pathogenic fungi are broadly classified as biotrophic or necrotrophic.  
Biotrophic pathogens (e.g., \textit{Blumeria graminis}, \textit{Puccinia spp.}) extract nutrients through haustoria without killing host cells, often producing only subtle changes in optical response.  
Necrotrophic fungi (e.g., \textit{Alternaria alternata}, \textit{Botrytis cinerea}), on the other hand, directly penetrate and degrade host tissues, leading to pigment loss, cellular collapse, and melanin accumulation that increase visible-band absorption and suppress NIR scattering~\cite{mahlein2016plant}.
The present study focuses on this necrotrophic infection regime.  
Melanized hyphae were modeled penetrating the cuticle and epidermis, accompanied by localized structural degradation and enhanced absorption.  
This representation captures both the geometric disruption and the compositional changes associated with fungal invasion, thereby providing a first-principles optical model that links pathogen-induced microstructural alterations to measurable hyperspectral signatures indicative of early-stage necrotrophic disease.

\begin{figure}[ht]
\centering
\includegraphics[width=0.5\linewidth]{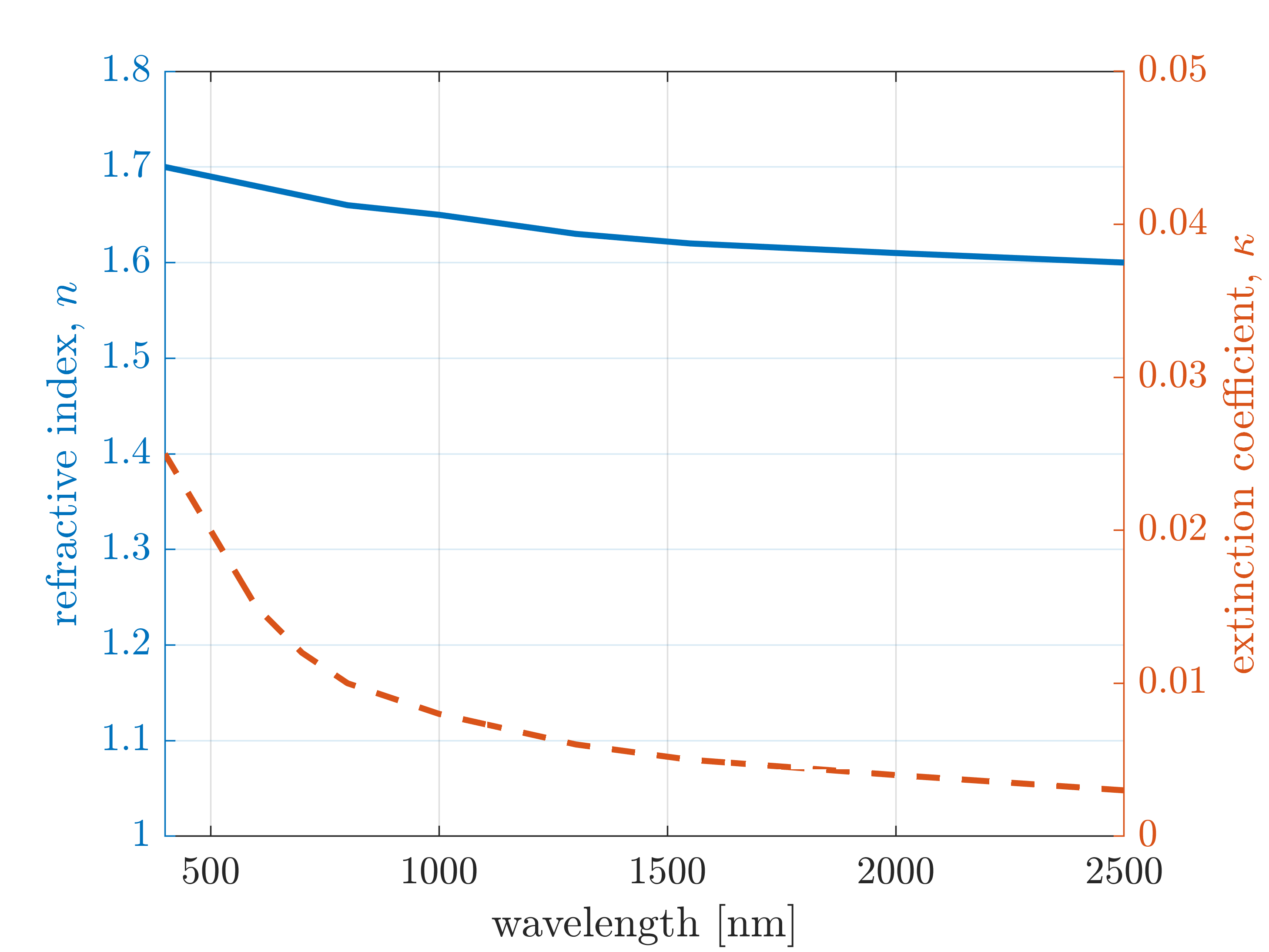}
\caption{Optical properties of the outer sheath of the fungal hyphae.}
\label{fig:melanin_optical_property}
\end{figure}

\begin{figure}[htbp]
    \centering
    \subfloat[\centering Diseased dicot (global view)]{
        \includegraphics[width=0.6\textwidth]{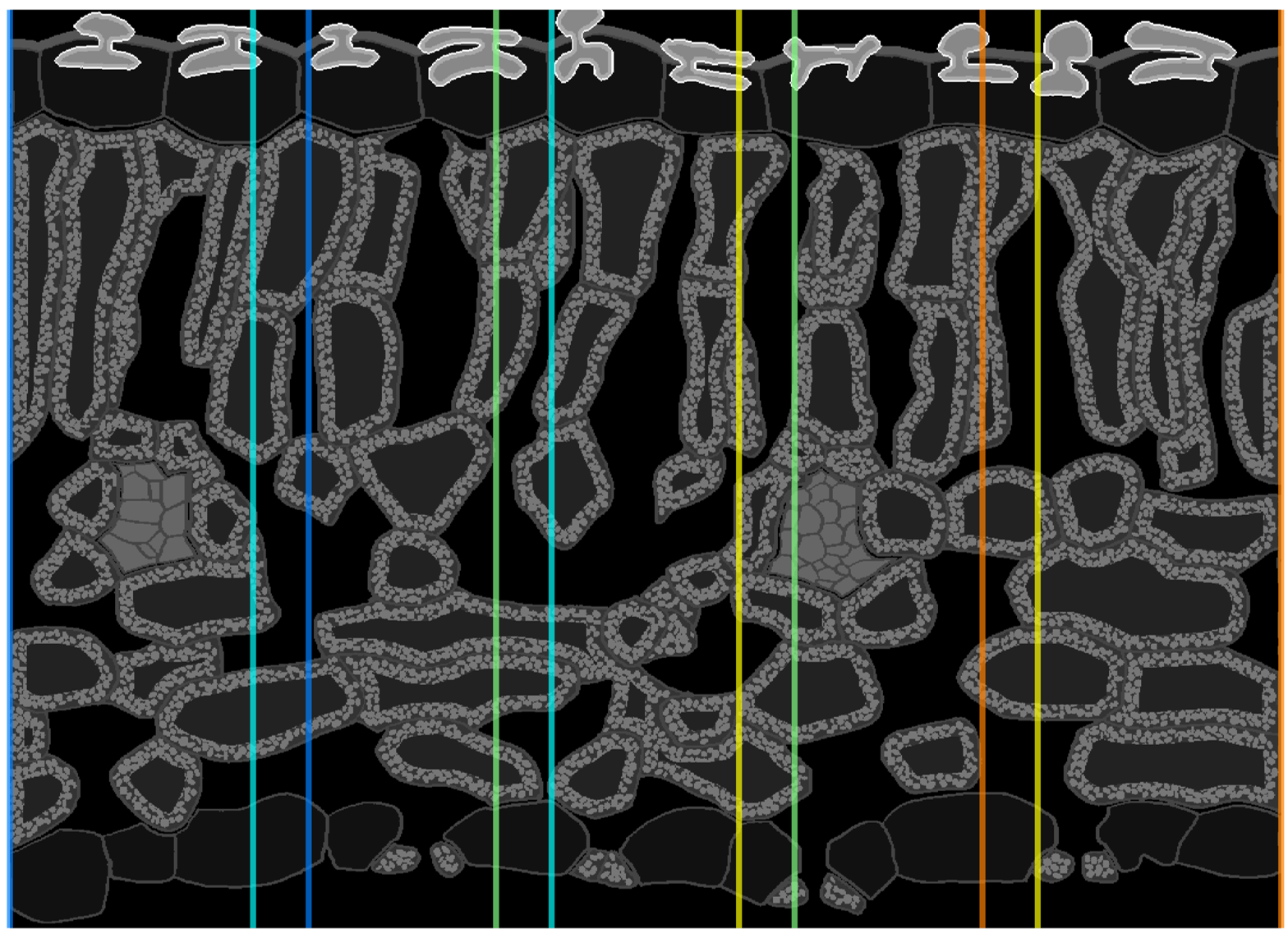}
        \label{fig:2D_FDTD_Yee_grid_global_D_sub1}
    }    
    \\
    \centering
    \subfloat[\centering Diseased dicot (slide view)]{
        \includegraphics[width=0.6\textwidth]{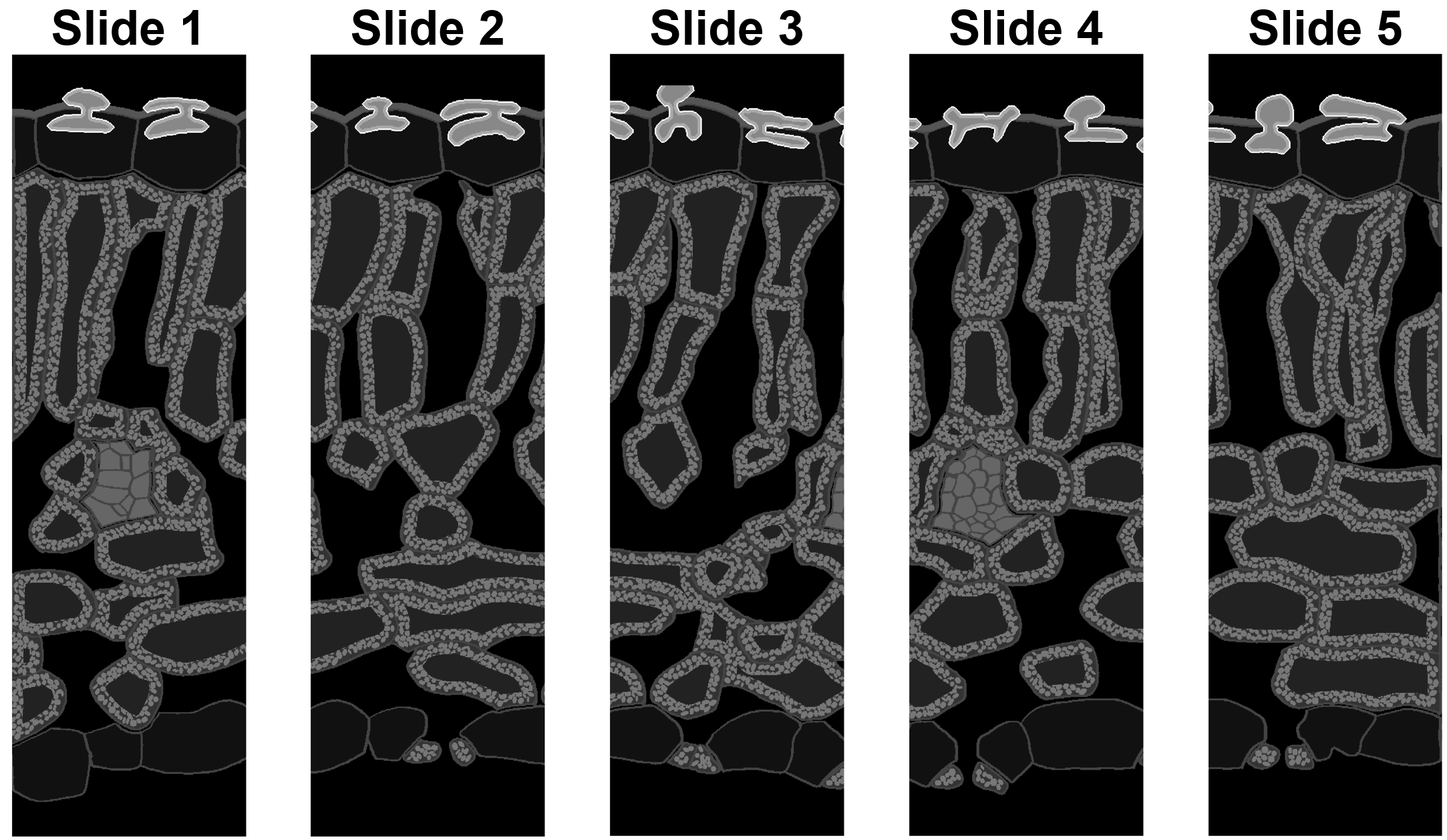}
        \label{fig:2D_FDTD_Yee_grid_global_D_sub2}
    }   
    \caption{Two-dimensional FDTD Yee grid for the cross-section of a dicot plant leaf with fungal mycelium, where differently colored regions indicate segmented tissues and fungal structures slightly penetrating the cuticle and epidermal layers on the upper surface of the leaf.}
    \label{fig:2D_FDTD_Yee_grid_global_D}
\end{figure}

The geometric model of the fungal-infected dicot plant leaf at an early stage is illustrated in Fig.~\ref{fig:2D_FDTD_Yee_grid_global_D}.
Since the goal of this study is a proof-of-concept simulation to observe the variation in spectral reflectance and transmittance depending on the presence of hyphae, the fungal structure was simply modeled as consisting of an inner cytoplasmic region surrounded by two layers-an inner and an outer sheath.
The inner sheath, in direct contact with the cytoplasm, was assumed to be composed of chitin, whereas the outer sheath was modeled as a composite layer of chitin and melanin.
Chitin, being a primary component of the cell wall, was assigned the same refractive index ($n = 1.52$) used previously for the cell wall modeling.
The outer sheath composed of chitin and melanin was modeled to incorporate the strong absorption characteristics of melanin, as shown in Fig.~\ref{fig:melanin_optical_property}.

\begin{figure}[ht]
\centering
\includegraphics[width=0.7\linewidth]{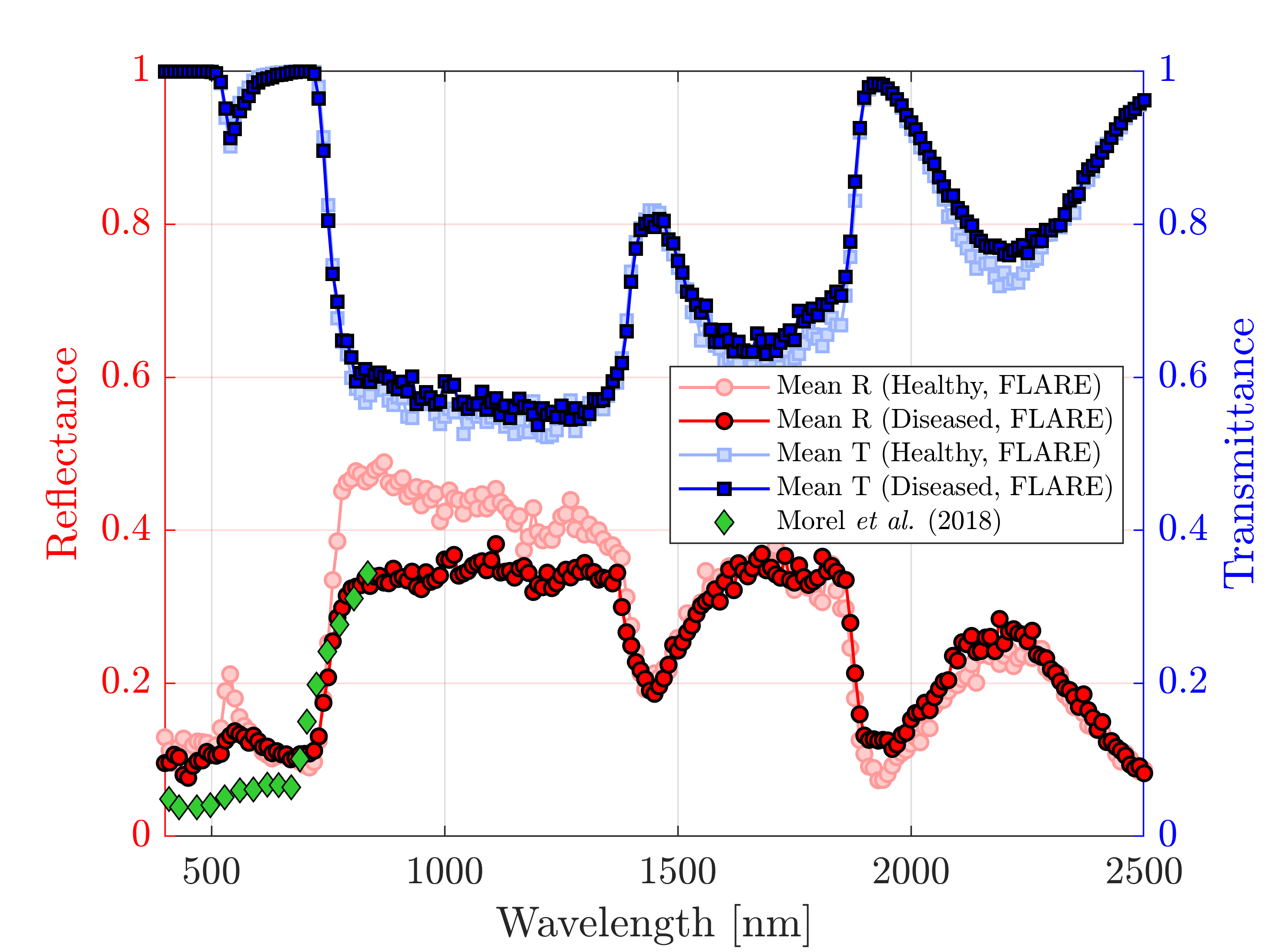}
\caption{Spectral reflectance and transmittance obtained from the FDTD simulation (averaged in terms of slides and  compared with measurement data by \cite{morel2018exploring}.}
\label{fig:validation_D_result}
\end{figure}

Fig. \ref{fig:validation_D_result} shows the comparison of the spectral reflectance and transmittance for healthy and diseased dicot sample plat leaf images.
Again, we ran several FDTD simulations in terms of wavelength, polarization, and sections, respectively, and average them at each wavelength, denoted as mean reflectance and transmittance in the figure.
The observations from Fig. \ref{fig:validation_D_result} can be summarized as follows.
First, the transmittance of the plant leaf shows no significant difference between the healthy and diseased cases. 
However, a slight decrease is observed across the entire spectrum, which is attributed to the broadband absorption characteristics of melanin pigments.
Second, a noticeable difference appears in the reflectance. 
In the VIS wavelength region, the healthy leaf exhibits a high reflectance peak in the green-light region, whereas the diseased leaf shows a significantly reduced green reflectance peak. 
This reduction is also attributed to the melanin pigments, which absorb light across the entire visible spectrum regardless of wavelength.
In addition, the abrupt increase in reflectance around 700 nm becomes much more gradual, and the reflectance in the NIR region (700-1,400 nm) decreases markedly. 
This trend of reduced NIR reflectance is consistent with previous studies that reported similar behavior in diseased plant leaves.
We compared our simulation results with experimental data in the VIS and NIR regions \cite{morel2018exploring}.
It can be observed that the overall trends described above are present in both cases, except that the experimental reflectance in the VIS wavelengths is slightly lower than that obtained from the FDTD simulations.
We expect that this discrepancy could be mitigated by adopting more realistic and advanced geometric and material modeling.

\begin{figure}[htbp]
\centering
\includegraphics[width=1\linewidth]{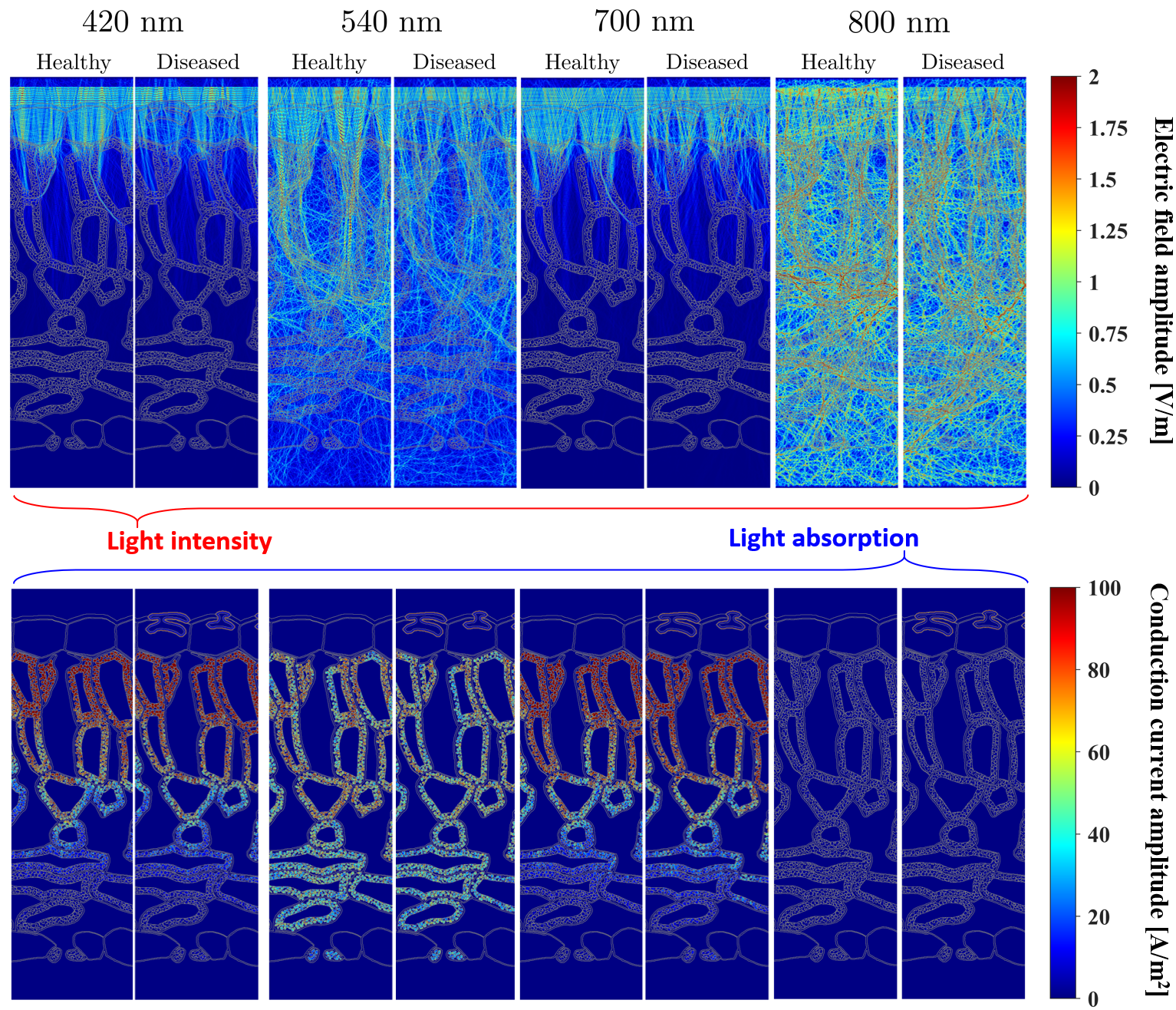}
\caption{Comparison of electric field amplitude and conduction current density distributions in the dicot healthy and diseased leaves (slide 2) at different wavelengths.}
\label{fig:Dicot_H_D_field_healthy}
\end{figure}
The physical origin of the reduced reflectance in diseased leaves can be understood by examining the field distributions.  
Fig.~\ref{fig:Dicot_H_D_field_healthy} compares the electric-field intensity and conduction-current maps at four representative wavelengths (420, 540, 700, and 800~nm) for healthy and diseased dicot samples.
At 420~nm and 700~nm, the healthy leaf exhibits strong epidermal lensing, which concentrates light into the mesophyll and produces large conduction currents indicative of efficient pigment absorption.  
In contrast, the diseased leaf shows pronounced absorption within melanized hyphae penetrating the epidermis, reducing the energy delivered to the mesophyll and degrading the focusing effect.
At 540~nm, the healthy sample shows weaker mesophyll absorption and stronger internal scattering.  
However, in the diseased case, the high refractive index and absorption of the hyphae increase surface reflectance and further suppress epidermal lensing, lowering the internal field amplitude.
A similar trend occurs at 800~nm: enhanced pre-entry reflection and attenuated focusing in the diseased leaf lead to reduced penetration depth compared with the healthy counterpart.
Overall, the FDTD results clearly show that melanin-rich fungal structures disrupt epidermal lensing and introduce strong localized absorption, thereby altering internal field distributions and reducing effective light delivery to deeper tissues.  
These full-wave simulations provide a physics-based interpretation of hyperspectral signatures associated with early-stage necrotrophic infection.

\section{Conclusion and Future Works}
This study introduced a full-wave optical modeling framework, based on the Finite-Difference Time-Domain (FDTD) method, for simulating light scattering and absorption in anatomically realistic plant leaves.  
By assigning wavelength-dependent complex refractive indices to digitally segmented tissue structures, the proposed approach accurately reproduced the characteristic visible–to–near-infrared reflectance and transmittance spectra of healthy monocot and dicot leaves, exhibiting excellent agreement with PROSPECT-PRO predictions.  
The use of CUDA-based GPU acceleration enabled large-scale simulations---exceeding tens of millions of grid cells---to be executed within practical runtimes, making full-wave modeling feasible for leaf-scale optical problems.  
Furthermore, by incorporating melanized hyphae and local structural deformation associated with early necrotrophic infection, the model successfully predicted measurable optical signatures arising from microscale pathological changes.

Future work will extend this framework to fully three-dimensional geometries, incorporate stochastic anatomical variability, integrate more advanced dispersive material models, and explore inverse-modeling approaches for quantitative leaf-physiology retrieval and early disease diagnostics using hyperspectral measurements.  
A deeper understanding of light dynamics within leaf tissues is expected to provide new physical insights into how microscale anatomical and biochemical changes manifest as measurable optical signatures.  
Such insights can support earlier and more reliable disease detection, improve the effectiveness and efficiency of risk mitigation and disease management strategies, and inform evidence-based policies for the optimized use of chemical inputs (e.g., fungicides, pesticides, and fertilizers) in plant disease control and surveillance of emerging plant pathogens.

Future research will extend this work in several directions:
\begin{itemize}
\item Three-dimensional volumetric modeling will enable more accurate representation of stomatal chambers, vascular networks, and mesophyll topology.  
\item Time-evolving simulations of disease progression may clarify how microstructural degradation produces spectral shifts over the infection cycle.  
\item Integration with inversion or learning-based retrieval algorithms could allow estimation of mesophyll structure or infection severity directly from measured spectra.  
\item Incorporating polarization and chlorophyll fluorescence will broaden applicability to BRDF polarimetry and solar-induced fluorescence studies.  
\item Coupling leaf-scale FDTD with canopy-level radiative-transfer models may establish a multiscale pathway linking cellular anatomy to airborne or satellite hyperspectral observations.
\end{itemize}

Overall, the proposed full-wave framework provides a rigorous physics-based foundation for studying light--leaf interactions and offers new opportunities for hyperspectral plant-disease detection and precision agriculture.

\section*{Acknowledgments}
This work was supported by the National Research Foundation of Korea (NRF) ...

\bibliographystyle{plain}
\bibliography{FDTD_bio}

\end{document}